%% file: TOP-19-008_temp.tex
\pdfoutput=1

\documentclass[11pt,twoside,a4paper,cmspaper,final,collab]{cms-tdr}
\def\svnVersion{f17197d-D}\def\svnDate{2020/11/09}\def\cmsCernNoTag{CERN-EP-2020-152}\def\cmsCernDate{\today}\def\cmsMessage{"Published in Physical Review D as \href{http://dx.doi.org/10.1103/PhysRevD.102.092013}{\doi{10.1103/PhysRevD.102.092013}.}"}

\begin{document}\cmsNoteHeader{TOP-19-008}

\hyphenation{had-ron-i-za-tion}
\hyphenation{cal-or-i-me-ter}
\hyphenation{de-vices}
\newlength\cmsFigWidth
\ifthenelse{\boolean{cms@external}}{\setlength\cmsFigWidth{0.49\textwidth}}{\setlength\cmsFigWidth{0.6\textwidth}}
\ifthenelse{\boolean{cms@external}}{\renewcommand{\CL}{\ensuremath{\text{C.L.}}\xspace}}{}
\ifthenelse{\boolean{cms@external}}{\providecommand{\CLnp}{\ensuremath{\text{C.L}}\xspace}}{\providecommand{\CLnp}{\ensuremath{\text{CL}}\xspace}}

\newlength\cmsTabSkip\setlength{\cmsTabSkip}{1ex}

\newcommand{\mur}{\ensuremath{\mu_\mathrm{R}}\xspace}
\newcommand{\muf}{\ensuremath{\mu_\mathrm{F}}\xspace}

\newcommand{\Mbl}{\ensuremath{M_{{\PQb}{\PQb}{\ell}{\ell}}}\xspace}
\newcommand{\yt}{\ensuremath{Y_{\PQt}}\xspace}
\newcommand{\Dytt}{\ensuremath{\Delta y_{\ttbar}}\xspace}
\newcommand{\Mtt}{\ensuremath{M_{\ttbar}}\xspace}
\newcommand{\Dybl}{\ensuremath{\Delta y_{{\PQb}\ell {\PQb}\ell}}\xspace}
\newcommand{\absDybl}{\ensuremath{\abs{\Delta y_{{\PQb}\ell {\PQb}\ell}}}\xspace}

\newcommand{\ytbestfit}{\ensuremath{\yt=1.16}\xspace}
\newcommand{\ytresult}{\ensuremath{\ytbestfit^{+0.24}_{-0.35}}\xspace}
\newcommand{\ytresultfull}{\ensuremath{\ytbestfit^{+0.07}_{-0.08}\stat^{+0.23}_{-0.34}\syst}\xspace}
\newcommand{\twosigmalim}{1.54}
\providecommand{\TOPpp} {{\textsc{Top++}}\xspace}
\hyphenation{multi-differential}

\title{Measurement of the top quark Yukawa coupling from \texorpdfstring{\ttbar}{ttbar} kinematic distributions in the dilepton final state in proton-proton collisions at \texorpdfstring{$\sqrt{s}=13\TeV$}{sqrt(s)=13 TeV}}

\date{\today}

\abstract{A measurement of the Higgs boson Yukawa coupling to the top quark is presented using proton-proton collision data at $\sqrt{s}=13\TeV$, corresponding to an integrated luminosity of 137\fbinv, recorded with the CMS detector. The coupling strength with respect to the standard model value, $\yt$, is determined from kinematic distributions in \ttbar final states containing ${\Pe}{\Pe}$, ${\Pgm}{\Pgm}$, or ${\Pe}{\Pgm}$ pairs. Variations of the Yukawa coupling strength lead to modified distributions for \ttbar production. In particular, the distributions of the mass of the \ttbar system and the rapidity difference of the top quark and antiquark are sensitive to the value of $\yt$. The measurement yields a best fit value of \ytresult, bounding $\yt<\twosigmalim$ at a 95\% confidence level. 
}

\hypersetup{
pdfauthor={CMS Collaboration},
pdftitle={Measurement of the top quark Yukawa coupling from tt kinematic distributions in the dilepton final state in proton-proton collisions at sqrt(s)=13 TeV},
pdfsubject={CMS},
pdfkeywords={CMS, top quark pairs, Yukawa coupling}
}
\maketitle

\section{Introduction}
\label{S:intro}

Since the discovery of the Higgs boson in 2012~\cite{higgsATLAS,Chatrchyan:2012xdj_higgsCMSlong}, one of the main goals of the CERN LHC program has been to study in detail the properties of this new particle.  In the standard model (SM), all fermions acquire their mass through the interaction with the Higgs field. More specifically, the mass of a given fermion, $m_\mathrm{f}$, arises from a Yukawa interaction with coupling strength $g_\mathrm{f}=\sqrt{2}m_\mathrm{f}/v$, where $v$ is the vacuum expectation value of the Higgs field. Among all such couplings, the top quark Yukawa coupling  is of particular interest.  It is not only the largest, but also remarkably close to unity. Given the measured top quark mass~\cite{topmassATLAS,Sirunyan:2018gqx_topmass18}, the mass-Yukawa coupling relation implies a value of the Yukawa coupling $g_{\PQt}^\mathrm{SM} \approx 0.99$ when evaluated near the energy scale of $m_{\PQt}$. Physics beyond the SM, such as two Higgs doublet and composite Higgs boson models, introduce modified couplings that alter the interaction between the top quark and the Higgs field~\cite{twoHDM,minHcomp}. This makes the interaction of the Higgs boson with the top quark one of the most interesting features of the Higgs field to study at the LHC today, especially because it is experimentally accessible through multiple avenues, both direct and indirect.

For the purpose of this measurement, we define for the top quark the parameter 
$\yt = g_{\PQt}/g_{\PQt}^\mathrm{SM}$, which is equivalent to the modifier $\kappa_{\PQt}$ introduced in the $\kappa$-framework~\cite{kappaFramework}. We consider only the case where $\yt\geq 0$, though certain specific techniques are sensitive also to the sign of the Yukawa coupling (for example, Ref.~\cite{Sirunyan:2018lzm_THQ}). Recent efforts have had notable success in directly probing $g_{\PQt}$ via the production of a Higgs boson in association with a top quark pair ({\ttbar}{\PH})~\cite{ttH, tthATLAS}. Currently, the most precise determination comes from the $\kappa$-framework fit in Ref.~\cite{combinedYukawa}, which yields $\yt=0.98\pm 0.14$ by combining information from several Higgs boson production and decay channels. These measurements, however, fold in assumptions of the SM branching fractions via Higgs couplings to other particles. Another way to constrain $g_{\PQt}$, which does not depend on these couplings, was presented in the search for four top quark production in Ref.~\cite{fourtopRun2}, yielding a limit of $\yt<1.7$ at a 95\% confidence level (\CL). However, it is also possible to constrain $g_{\PQt}$ indirectly using the kinematic distributions of reconstructed \ttbar pair events, a technique that has been recently used by CMS to derive a similar limit of $\yt<1.67$ at 95\% \CL in the lepton+jets \ttbar decay channel~\cite{ytpaper}. The measurement presented in this paper follows this last approach, but in the dilepton final state.

\begin{figure*}[htb]
    \centering
    \includegraphics[width=.35\textwidth]{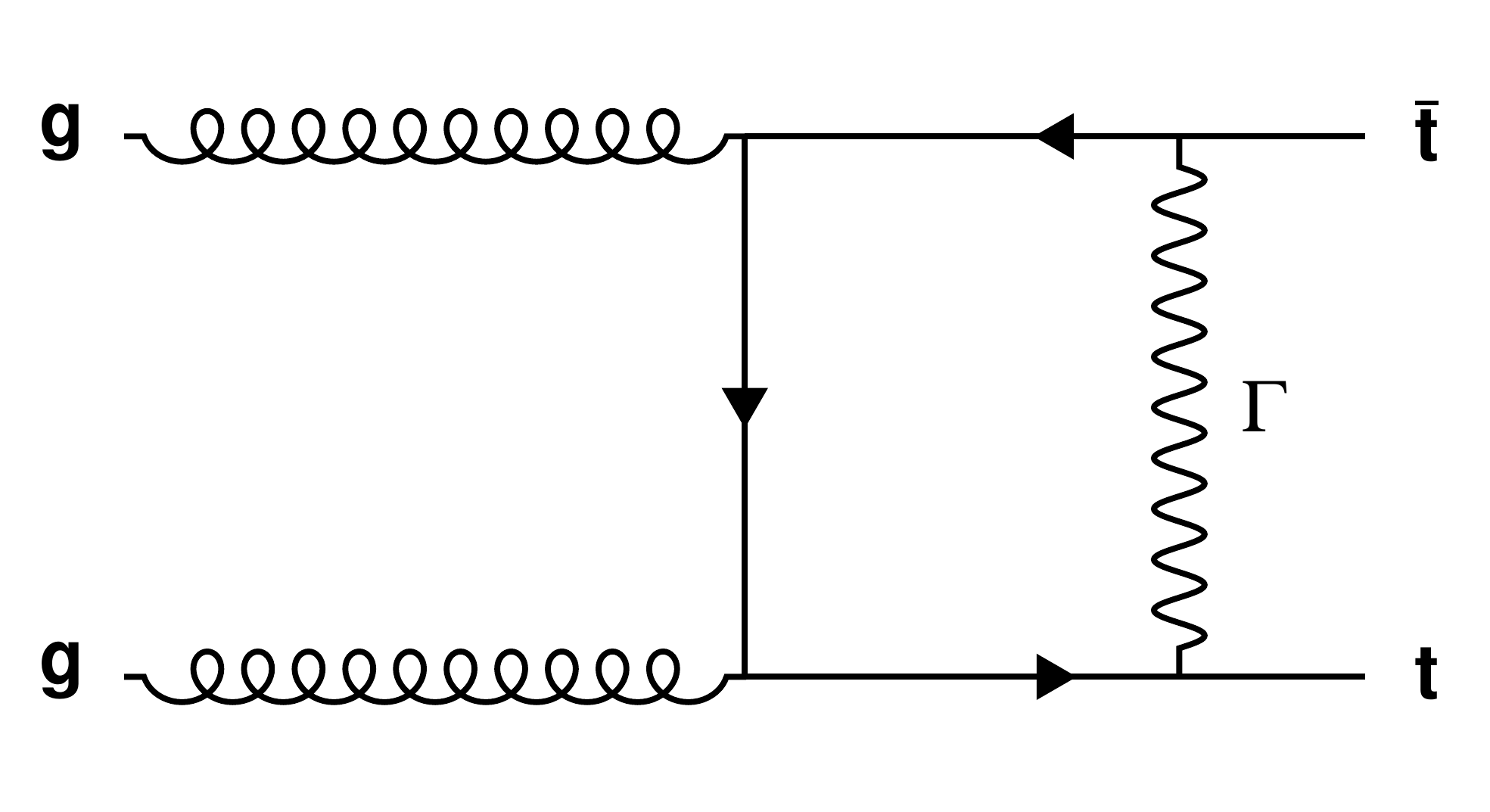}\quad\quad\quad\includegraphics[width=.35\linewidth]{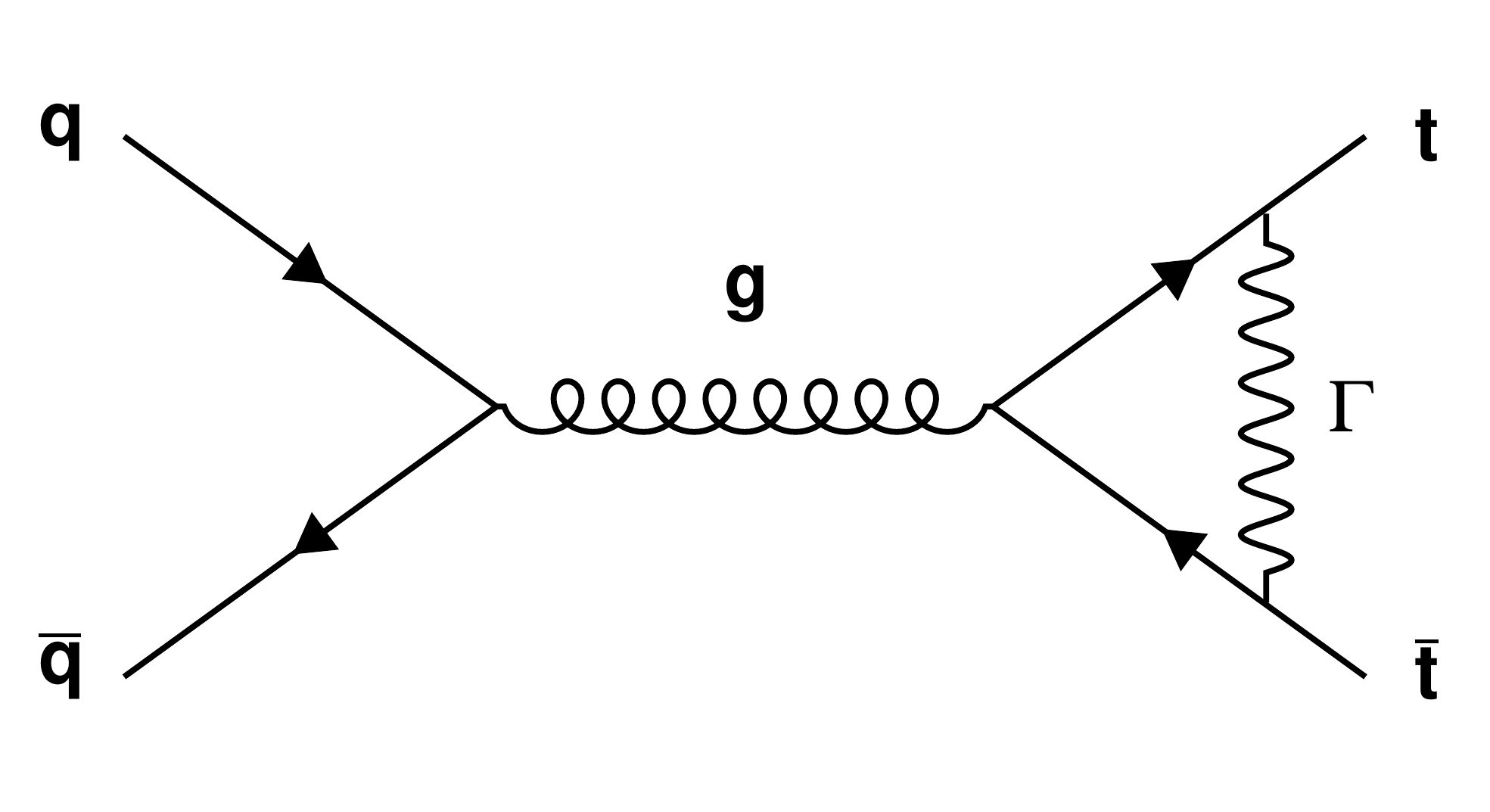}
    \caption{Sample Feynman diagrams for EW contributions to gluon-induced and quark-induced top quark pair production, where $\Gamma$ stands for neutral vector and scalar bosons.}
    \label{fig:diagrams}
\end{figure*}

\begin{figure*}[htb]
    \centering
    \includegraphics[width=.48\textwidth]{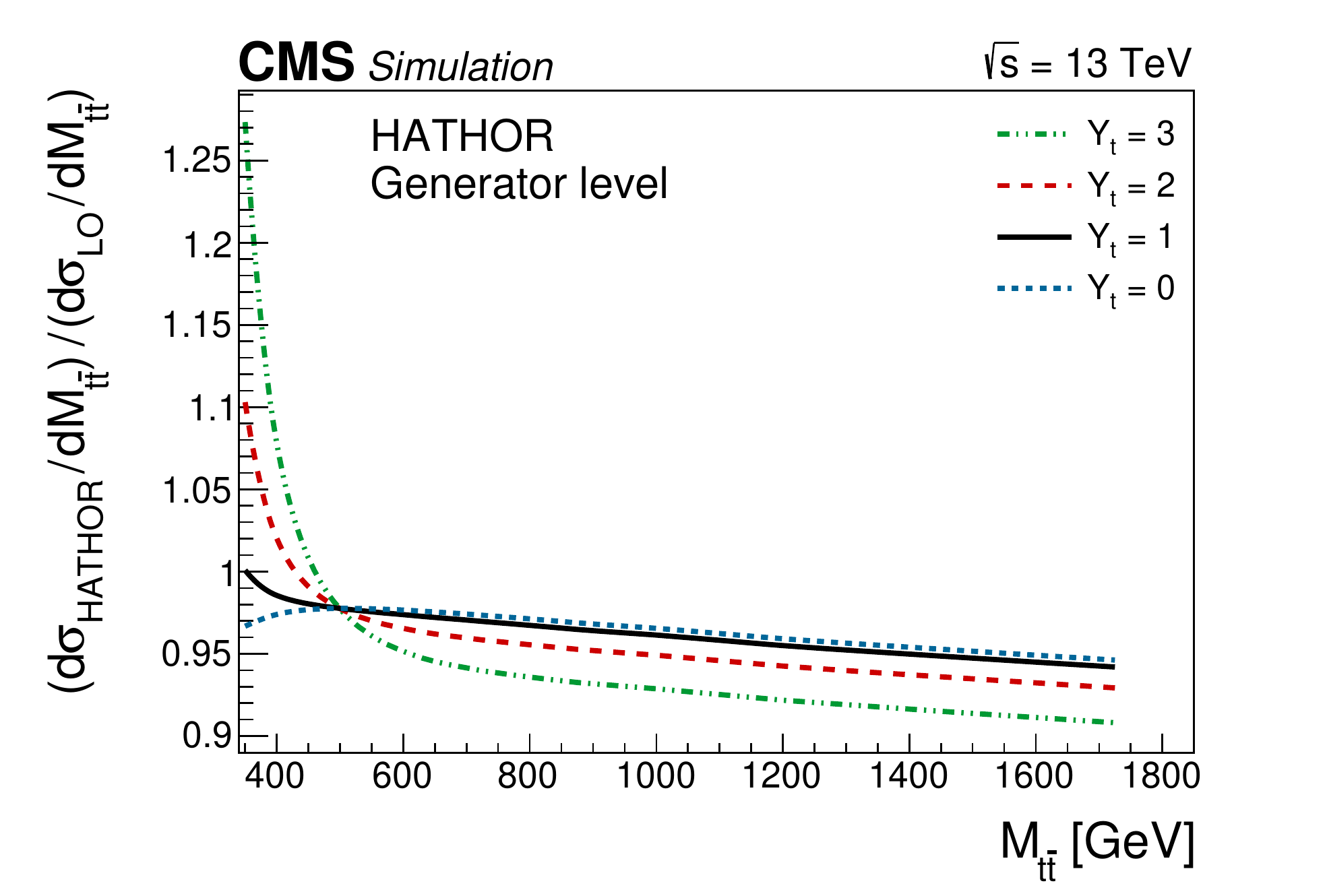}
    \includegraphics[width=.48\textwidth]{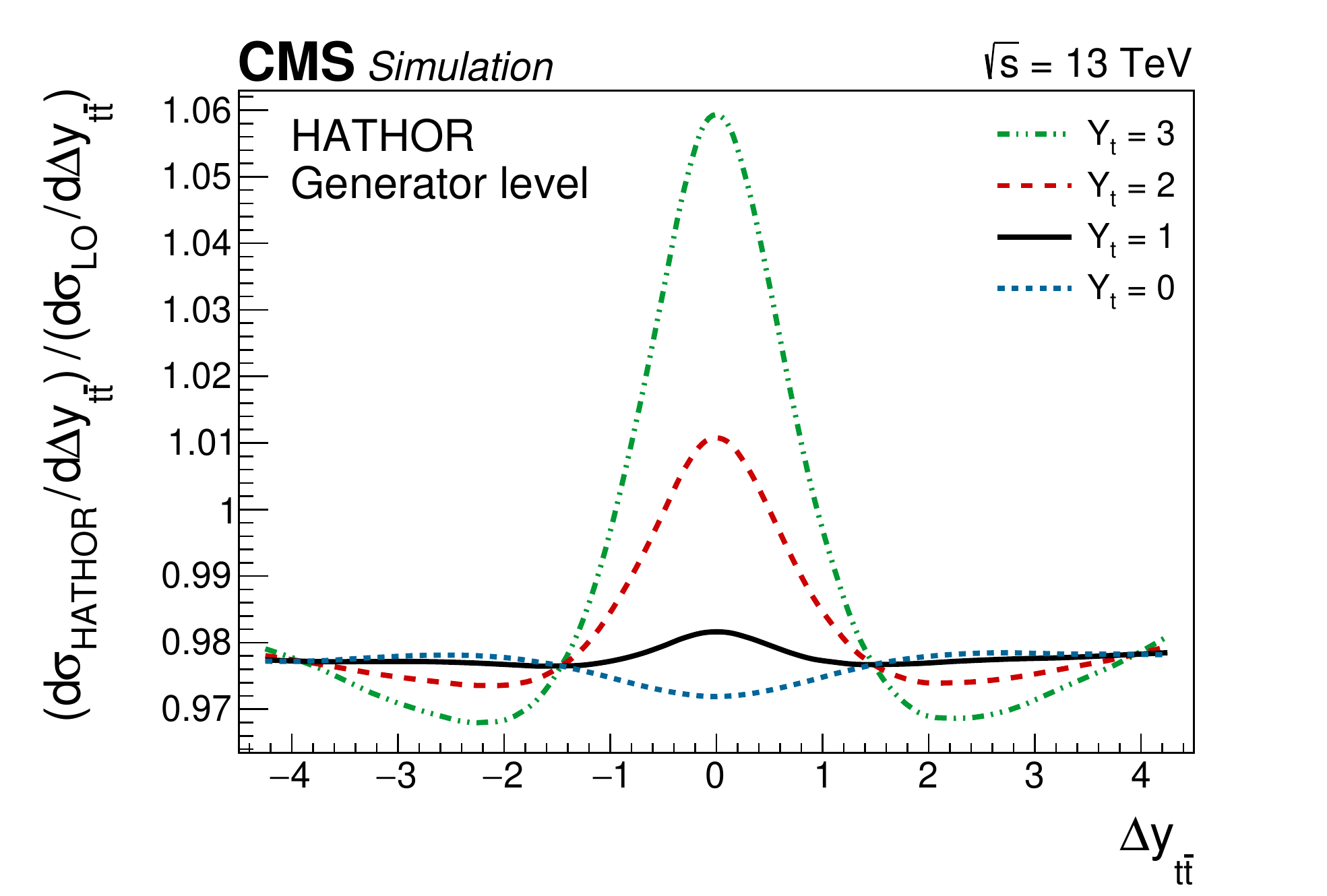}
    \caption{Effect of the EW corrections on \ttbar differential kinematic distributions for different values of $\yt$, after reweighting of simulated events. The effect is shown on the distribution of the invariant mass, \Mtt (left), and the difference in rapidity between the top quark and antiquark, $\Delta y_{\ttbar}$ (right).}
    \label{fig:hatratgen}
\end{figure*}

Current commonly used Monte Carlo (MC) simulations of \ttbar production include next-to-leading-order (NLO) precision in perturbative quantum chromodynamics (QCD). Subleading-order corrections arise from including electroweak (EW) terms in the perturbative expansion of the strong coupling $\alpS$ and the EW coupling $\alpha$. Such terms begin to noticeably affect the cross section only at loop-induced order $\alpS^2 \alpha$, and are typically not included in current MC simulation. While these terms have a very small effect on the total cross section, they can alter the shape of kinematic distributions to a measurable extent. Such changes become more noticeable if the Yukawa coupling affecting the loop correction (Fig.~\ref{fig:diagrams}) is anomalously large. Therefore, these corrections are of particular interest in deriving upper limits on $g_{\PQt}$. For example, the distribution of the invariant mass of the \ttbar system, \Mtt, will be affected significantly by varying \yt. Doubling the value of \yt can alter the \Mtt distribution by about 9\% near the \ttbar production threshold, as described in Ref.~\cite{UwerWeak}.
Another variable sensitive to the value of \yt is the difference in rapidity between the top quark and antiquark, $\Delta y_{\ttbar} = y({\PQt})-y(\overline{\PQt})$. In \ttbar production, \Mtt and $\Delta y_{\ttbar}$ are proxies for the Mandelstam kinematic variables $s$ and $t$, respectively, which span the event phase space and can thus be used to include the EW corrections in previously generated event samples via reweighting. The effects of these corrections are shown for differential cross sections of \Mtt and $\Delta y_{\ttbar}$ in Fig.~\ref{fig:hatratgen}. These are computed by reweighting simulated \ttbar events at generator level using predictions from the \textsc{hathor} software package~\cite{hathorart}.

 After calculating the dependence of these corrections on \yt, a measurement is performed. We use events in the dilepton final state (${\Pe}{\Pe}$, ${\Pgm}{\Pgm}$, or ${\Pe}{\Pgm}$), for which this type of measurement has not yet been performed. While this decay channel has a smaller branching fraction than the lepton+jets channel studied in Ref.~\cite{ytpaper}, it has lower backgrounds due to the  presence of two final-state high-\pt leptons. However, two neutrinos are also expected in this final state, which escape detection and pose challenges in the kinematic reconstruction. For this reason, we do not perform a full kinematic reconstruction as was done in the previous measurement in the lepton+jets channel. This measurement also utilizes a much larger data set with an integrated luminosity of 137\fbinv collected during Run 2 at the LHC from 2016 to 2018, allowing us to achieve comparable precision to that in Ref.~\cite{ytpaper} for a decay channel with a much lower branching fraction.

In this paper, we will first briefly describe the CMS detector (Section~\ref{sec:detector}), and then discuss the data and MC samples (Section~\ref{S:simulation}), followed by the methods for event selection (Section~\ref{S:selection}) and reconstruction (Section~\ref{S:reco}). We then present an outline of the measurement technique (Section~\ref{S:stats}) and the contributing sources of uncertainty (Section~\ref{S:sys}), and conclude with the results of the measurement (Section~\ref{S:fit}) and the summary (Section~\ref{S:conc}).

\section{The CMS detector}
\label{sec:detector}
The central feature of the CMS detector is a superconducting solenoid of 6\unit{m} internal diameter,
providing a magnetic field of 3.8\unit{T}. Within the solenoid volume are a silicon pixel and strip
tracker, a lead tungstate crystal electromagnetic calorimeter (ECAL), and a brass and scintillator
hadron calorimeter (HCAL), each composed of a barrel and two endcap sections. Forward calorimeters
extend the coverage provided by the barrel and endcap detectors. Muons
are measured in gas-ionization detectors embedded in the steel flux-return yoke outside the
solenoid. 

The particle-flow (PF) algorithm~\cite{ref:particleflow} aims to reconstruct and identify each individual particle in an event, with an optimized combination of information from the various elements of the CMS detector. The energy of photons is obtained from the ECAL measurement. The energy of electrons is determined from a combination of the electron momentum at the primary interaction vertex as determined by the tracker, the energy of the corresponding ECAL cluster, and the energy sum of all bremsstrahlung photons spatially compatible with originating from the electron track. The energy of muons is obtained from the curvature of the corresponding track. The energy of charged hadrons is determined from a combination of their momentum measured in the tracker and the matching ECAL and HCAL energy deposits, corrected for zero-suppression effects and for the response function of the calorimeters to hadronic showers. Finally, the energy of neutral hadrons is obtained from the corresponding corrected ECAL and HCAL energies. 

Events of interest are selected using a two-tiered trigger system~\cite{Khachatryan:2016bia}. The first level (L1), composed of custom hardware processors, uses information from the calorimeters and muon detectors to select events at a rate of around 100\unit{kHz} within a time interval of less than 4\mus. The second level, known as the high-level trigger, consists of a farm of processors running a version of the full event reconstruction software optimized for fast processing, and reduces the event rate to around 1\unit{kHz} before data storage.

A more detailed description of the CMS detector, together with a definition of the coordinate system and relevant kinematical variables, can be found in Ref.~\cite{Chatrchyan:2008zzk}.

\section{Simulation of top quark pair production and backgrounds}

\label{S:simulation}

The production of \ttbar events is simulated at the matrix-element (ME) level with NLO QCD precision, using the \POWHEG 2.0 (hvq) generator~\cite{Nason:2004rx,Frixione:2007vw,Alioli:2010xd,Frixione:2007nw}. The calculation is performed with the renormalization and factorization scales, \mur and \muf, set to the transverse top quark mass, $\mT=\sqrt{\smash[b]{m_{\PQt}^2+\pt^2}}$, where \pt is the transverse momentum of the top quark and the quantity is evaluated in the \ttbar rest frame. The default value of $m_{\PQt}$ is set to 172.5\GeV. The ME calculations obtained from \POWHEG are combined with the parton shower simulation from \PYTHIA 8.219~\cite{Sjostrand:2014zea}, using the underlying-event tune M2T4~\cite{ISR_FSR} to simulate data taken in 2016, and \PYTHIA 8.226 using the tune CP5~\cite{CP5} to simulate data taken in 2017 and 2018. The parton distribution function (PDF) set NNPDF3.0 at NLO~\cite{NNPDF} is used for 2016 and updated to {NNPDF}3.1~\cite{NNPDF31} at next-to-NLO (NNLO) for 2017 and 2018. These samples are normalized to a \ttbar cross section calculated at NNLO in QCD including resummation of next-to-next-to-leading logarithmic (NNLL) soft gluon terms with \textsc{Top++} 2.0~\cite{Czakon:2011xx}. The calculation uses the PDF4LHC prescription~\cite{Botje:2011sn__pdf4lhc} with the MSTW2008 NNLO~\cite{Martin:2009iq__9,Martin:2009bu__10}, CT10 NNLO~\cite{Lai:2010vv__11,Gao:2013xoa__12} and NNPDF2.3~\cite{Ball:2012cx} PDF sets used to generate an envelope of uncertainty with the midpoint of the envelope used for the central predictions. The PDF uncertainty is then summed in quadrature with the scale uncertainty to arrive at an overall uncertainty of $\approx$5\% on the nominal value of 832\unit{pb}. The shape effects associated with the PDF uncertainty are considered separately in Section~\ref{S:sys}.

A high purity of \ttbar events can be obtained in the dilepton channel, as shown in Section~\ref{S:selection}. A small contamination is expected to result from background processes, which are modeled by simulation. In particular, we account for dilepton production due to Drell--Yan type processes and single top quark production. Other SM processes, such as \PW boson production, were investigated and found to have negligible contributions. Diboson production is also included, although its expected contribution is minute due to the small total cross section of the process.

About 1\% of the events identified as \ttbar dilepton decays are misidentified \ttbar lepton+jets decays. EW corrections are applied to all \ttbar events,  even  misidentified ones, so their kinematic distributions remain dependent on \yt. Thus, these events are still considered as signal, even though their contribution to the measurement sensitivity is greatly diminished relative to dilepton events.

Single top quark events are simulated at NLO with \POWHEG in combination with \PYTHIA, while diboson events are simulated with \PYTHIA at leading-order (LO) QCD precision. Drell--Yan production is simulated at LO using \MGvATNLO version 2.2.2 for 2016 and version 2.2.4 for 2017 onwards~\cite{Alwall:2014hca}, with up to 4 additional partons, interfaced to \PYTHIA using the MLM matching algorithm~\cite{Mangano:2006rw_MLMorig,MLM}.

The detector response to all simulated events is modeled with the \GEANTfour software toolkit~\cite{GEANT4}.
In addition, the effects of multiple proton-proton interactions per event are included in simulations and the distribution of these pileup interactions is reweighted to the vertex multiplicity distribution in the data.

\subsection{Simulation of electroweak corrections}

Contributions to the top quark pair production arising from QCD+EW diagrams are evaluated using the \textsc{hathor} package~\cite{hathorart}, which is used to compute a double-differential cross section as a function of \Mtt and $\Delta y_{\ttbar}$ including LO QCD diagrams and certain EW diagrams of order $\alpS^2 \alpha$. These diagrams involve massive boson exchange and examples are shown in Fig.~\ref{fig:diagrams}. The contributions from photon-mediated interactions are not included. Contributions from diagrams involving virtual photon exchange should not be assessed individually, as they are partially cancelled not only by real emission diagrams but also by contributions from $\Pg\PGg$ production~\cite{ignorephotons}. A complete assessment would require the modeling of photon content within protons. This was not performed here, as the net effect is fairly small. For example, Ref.~\cite{ignorephotons} cites a 1\% effect from photon-mediated contributions to the \ttbar cross section at the LHC with detector-based kinematic cuts. Thus, we include only diagrams involving massive vector and scalar boson interactions, which are the dominant EW diagrams at this order.

The ratio of this double-differential cross section is evaluated with respect to the LO QCD computation, in order to obtain a multiplicative weight correction $w(\Mtt,\Delta y_{\ttbar})$. Applying this weight at parton level to MC samples produced at NLO QCD approximates the inclusion of EW corrections in the simulation. This \textit{multiplicative approach} to including EW corrections was used previously in Ref.~\cite{ytpaper}, and has the benefit of approximating the inclusion of diagrams at order $\mathcal{O}(\alpS^3\alpha)$. Because EW corrections factorize in some kinematic regimes, this is a better-motivated approach than the alternative \textit{additive approach}, in which one adds the fixed-order result at order $\mathcal{O}(\alpS^2\alpha)$ while ignoring all potential contributions of order $\mathcal{O}(\alpS^3\alpha)$ (see Ref.~\cite{Czakon:2017NLOEW} for a more detailed discussion). In other words, the additive approach applies the EW correction factor only to the proportion of \POWHEG events present at LO QCD, while any interplay between EW corrections and higher-order QCD simulation is ignored. Although the multiplicative approach is clearly favored, neither approach can account for the effects of two-loop contributions near the \ttbar production threshold. To account for the lack of knowledge of such terms, we take the difference between the two predictions as a modeling uncertainty in this regime, as suggested in Ref.~\cite{UwerWeak}. The estimation of this uncertainty is discussed further in Section~\ref{S:stats}.

The EW correction weights  are calculated for discrete integer values of $\yt=0,1,\ldots,5.$ Since the dependence of the production rate on \yt is exactly quadratic, these discrete values are sufficient to parametrize event yields as a continuous function of \yt (as discussed in Section~\ref{S:stats}). This allows us to measure which value of \yt best describes the data.

\section{Event and object selection}
\label{S:selection}

Events are selected using single-electron or single-muon triggers. Data taking at the LHC was interrupted by technical stops at the end of each year, leading to some changes in configuration and modeling between 2016, 2017, and 2018. For events selected by the single-electron trigger, we require a trigger \pt threshold of 27\GeV with the exception of 2018, where a threshold of 32\GeV is used. In the case of the single-muon trigger, we select events with a trigger \pt threshold of 24\GeV, which is raised to 27\GeV only for 2017 due to high event rates.

We ensure that all electrons and muons are within the silicon tracker coverage by requiring a pseudorapidity $\abs{\eta}<2.4$. To operate well above the trigger threshold, we then require at least one isolated electron or muon reconstructed with $\pt>30$\GeV, except in 2018, where we require leading \pt electrons to have $\pt>34$\GeV in accordance with the trigger threshold. The same lepton isolation criteria described in Ref.~\cite{Otto:17002} are used. After selecting the leading \pt lepton, a second isolated electron or muon with $\pt>20\GeV$ is required. Events with three or more isolated leptons with $\pt >15\GeV$ are discarded.

Jets are clustered from PF objects via the anti-\kt algorithm~\cite{antikt,fastjet} with a distance parameter of 0.4. The jet momentum is calculated as the vectorial sum of the momenta of its constituents. Corrections to the jet energy are derived as a function of jet \pt and $\eta$ in simulation and improved by measurements of energy balance in data~\cite{JES}. We select jets with $\abs{\eta}<2.4$ and $\pt>30$\GeV. 

Jets originating from {\PQb} quarks are identified using the DeepCSV algorithm~\cite{btagRun2}. The algorithm provides three working points: loose, medium, and tight, in order of decreasing efficiency and increasing purity. The {\PQb} identification efficiencies (and light quark misidentification rates) are 84 (11)\%, 68 (1.1)\%, and 50 (0.1)\%, respectively. For an initial selection, we consider events with a minimum of two {\PQb} jet candidates passing the loose working point of the algorithm. When applied to simulated \ttbar dilepton decays, we find that this initial selection of {\PQb} jets  will correctly include both {\PQb} jets originating from top quark decays in 87\% of events. In around 9\% of simulated \ttbar dilepton events passing this initial selection, there are more than two jets passing the loose working point, leading to an ambiguity in jet assignment. If such events have exactly two jets passing a higher working point (medium or tight), then those two jets are considered the viable candidates for {\PQb} jets originating from a top quark decay, and the ambiguity is resolved without using kinematic properties of the event. The small fraction of events with more than two viable {\PQb} jet candidates, making up 4\% of the initially selected \ttbar dilepton events, are discarded. After this selection procedure, each event remaining in the sample has exactly two {\PQb} jet candidates, which together are correctly identified in 85\% of simulated \ttbar dilepton events.

In order to remove Drell--Yan background events in the ${\Pe}{\Pe}$ and ${\Pgm}{\Pgm}$ channels, we reject events in which the two leptons have an invariant mass below 50\GeV or within 10\GeV around the \PZ boson mass of 91.2\GeV.

The missing transverse momentum vector (\ptvecmiss), defined as the negative vector sum of all transverse momenta, is generally of large magnitude in dilepton decays because of the two undetected final-state neutrinos. To further aid in removing Drell--Yan events, we impose an additional selection requirement on the magnitude of the missing transverse momentum, requiring  $\ptmiss>30$\GeV in all events with ${\Pe}{\Pe}$ or ${\Pgm}{\Pgm}$ in the final state. 

The breakdown of expected signal and background yields, summed over the three channels (${\Pe}{\Pe}$, ${\Pgm}{\Pgm}$, ${\Pe}{\Pgm}$), is shown by year in Table~\ref{tab:eventcounts}. The Drell--Yan background is estimated to be about 2\%. Single top quark production accounts for roughly another 2\% of the estimated sample composition.

\begin{table*}[htbp]
\centering
\topcaption{Simulated signal, background, and data event yields for each of the three years and their combination. The rightmost column shows the fraction of each component relative to the total simulated sample yield across the full data set. The statistical uncertainty in the simulated event counts is given.}
\newcolumntype{x}{D{,}{\,\pm\,}{6.3}}
\begin{scotch}{l x x x x r}
Source & \multicolumn{1}{c}{2016 (36\fbinv)} & \multicolumn{1}{c}{2017 (41\fbinv)} & \multicolumn{1}{c}{2018 (60\fbinv)} & \multicolumn{1}{c}{All (137\fbinv)} & \% total MC \\
\hline
\ttbar   & 140\,830, 130 & 170\,550,100 & 259\,620, 150 & 571\,010, 220 & 96.2\% \\
Drell--Yan &    1920,  50 &    2690, 80  &    4960, 130 &    9840, 170 & 1.7\% \\
Single \PQt &    3020, 30 &    3520, 20  &    5830,  30 &  12\,370,  50 & 2.1\% \\
Diboson  &     140,  10 &     150, 10  &     250,  20 &     540,  20 & 0.1\% \\[\cmsTabSkip]
Total    & 145\,940, 150 & 177\,400, 120 & 270\,660, 200 & 593\,760, 280 & \\
Data & \multicolumn{1}{l}{144\,817} & \multicolumn{1}{l}{178\,088} & \multicolumn{1}{l}{264\,791} & \multicolumn{1}{l}{587\,696} & \\
\end{scotch}
\label{tab:eventcounts}
\end{table*} 

\section{Event reconstruction}
\label{S:reco}

The EW corrections are calculated based on \Mtt and $\Delta y_{\ttbar} $. However, to evaluate these quantities it is necessary to reconstruct the full kinematic properties of the \ttbar system, including the two undetected neutrinos. While it is possible to completely reconstruct the neutrino momenta in the on-shell approximation, such a reconstruction is highly sensitive to \ptmiss, which introduces large resolution effects and additional systematic uncertainties. We observe that using the proxy variables $\Mbl = M({\PQb}+\overline{{\PQb}}+\ell+\overline{\ell})$ and $\absDybl = \abs{y({\PQb}+\overline{\ell}) - y(\overline{{\PQb}}+\ell)}$, where $\ell$ represents a final-state electron or muon, results in a more precise measurement.

Unlike $\Mbl$, the accurate reconstruction of \absDybl requires that each of the two {\PQb} jets is matched to the correct lepton, i.e., both originating from the same top quark decay. In order to make this pairing, we utilize the information from the kinematic constraints governing the neutrino momenta.

If one assumes the top quarks and \PW bosons to be on-shell, the neutrino momenta are constrained by a set of quadratic equations arising from the conservation of four-momentum at each vertex. We refer to these kinematic equations, collectively, as the mass constraint. The mass constraint for each top quark decay results in a continuum of possible solutions for neutrino momenta, which geometrically can be presented as an intersection of ellipsoids in three-dimensional momentum-space~\cite{burt}. For certain values of input momenta of {\PQb} jets and leptons these ellipsoids do not intersect at all, such that the quadratic equations have no real solution. In these scenarios, the mass constraint cannot be satisfied.

In cases where the mass constraint can be satisfied, one could also constrain \ptmiss in the event to equal the \pt sum of the two undetected neutrinos. We call this the \ptmiss constraint. This constraint reduces the remaining solutions to a discrete set, containing either two or four possibilities that fully specify the momenta of both neutrinos. Similar to the case of the mass constraint, there are some values of the input parameters for which the \ptmiss constraint cannot be satisfied. 

When looking at simulated events where both {\PQb} jets are correctly reconstructed and paired, we find that the mass constraint can be satisfied in 96\% of all cases, while the mass and \ptmiss constraints can be simultaneously satisfied in 55\% of cases. In contrast, if the {\PQb} jets are correctly reconstructed but incorrectly paired to leptons, the mass constraint can be satisfied in only 23\% of cases, while both mass and \ptmiss constraints can be met in only 18\% of cases.

Pairings with no solution to the mass constraint are thus frequently incorrect. When the mass constraint can be satisfied, pairings with a solution to the \ptmiss constraint are more likely to be correct. This information is used as part of the pairing procedure, which has three steps.
\begin{enumerate}

    \item The mass constraint is checked for both possible pairings. If only one pairing is found to satisfy the mass constraint, that pairing is used. If both pairings fail to satisfy the mass constraint, the event is discarded. If both pairings satisfy the mass constraint, we check the \ptmiss constraint.
    \item If only one pairing allows for the \ptmiss constraint while the other does not, the pairing yielding an exact solution to the \ptmiss constraint is used.
    \item If the kinematic variables of the neutrinos do not suggest a clear pairing, the {\PQb} jets, ${\PQb}_1$ and ${\PQb}_2$, are paired with the leptons ($\ell$, $\overline{\ell}$) by minimizing the quantity $$ \Sigma_{1(2)} =\Delta R({\PQb}_{1(2)},\ell)+ \Delta R({\PQb}_{2(1)},\overline{\ell})$$ among the two possible pairings, where $\Delta R({\PQb},\ell)=\sqrt{\smash[b]{(\eta_{\PQb} -\eta_\ell)^2 + (\varphi_{\PQb}-\varphi_\ell)^2}}$ and $\varphi$ is the azimuthal angle in the transverse plane. 

\end{enumerate}
In simulation, this procedure discards 7\% of the signal sample, targeting events which generally involve an improperly assigned or misidentified {\PQb} jet (at a rate of 72\%). This raises the fraction of events that successfully identify both {\PQb} jets from a top quark decay to 89\% in simulation. After these steps, we obtain the correct {\PQb} jet pairing in 82\% of simulated dilepton \ttbar events for which both {\PQb} jets originating from top quark decays were correctly identified, and thus 73\% of simulated dilepton \ttbar events overall.

\begin{figure*}[bt!]
    \centering
    \includegraphics[width=.48\linewidth]{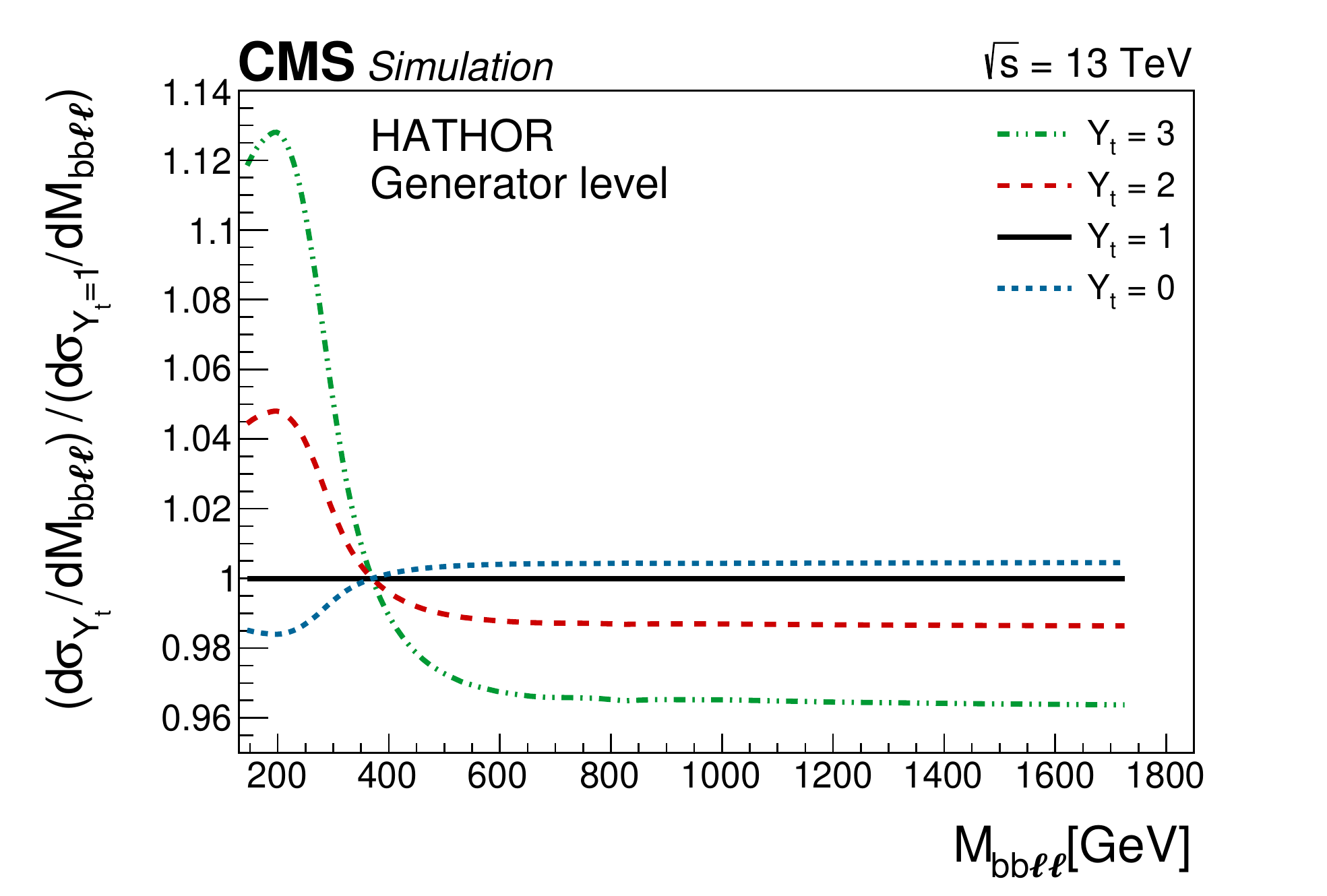}\;
    \includegraphics[width=.48\linewidth]{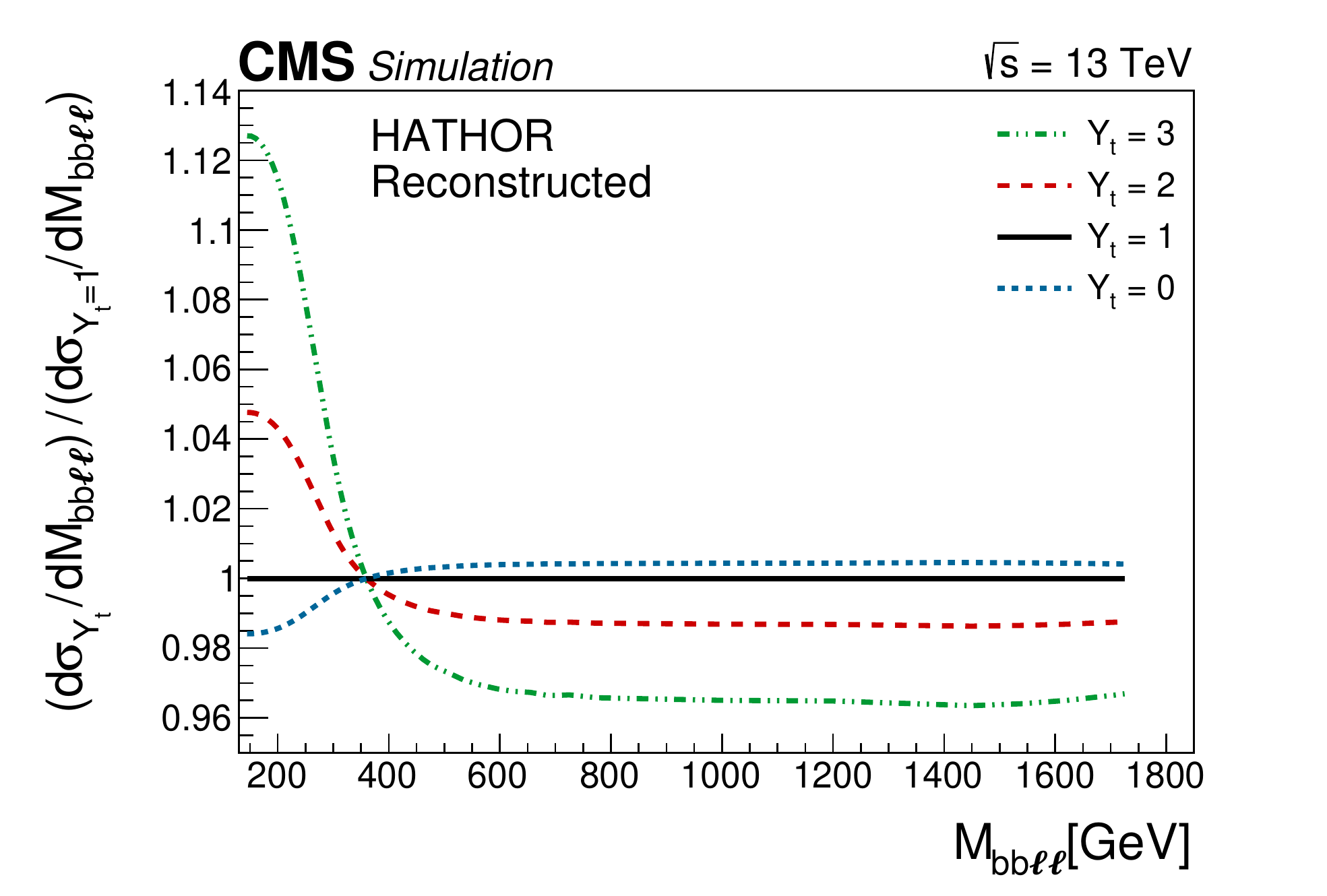}
  
    \includegraphics[width=.48\linewidth]{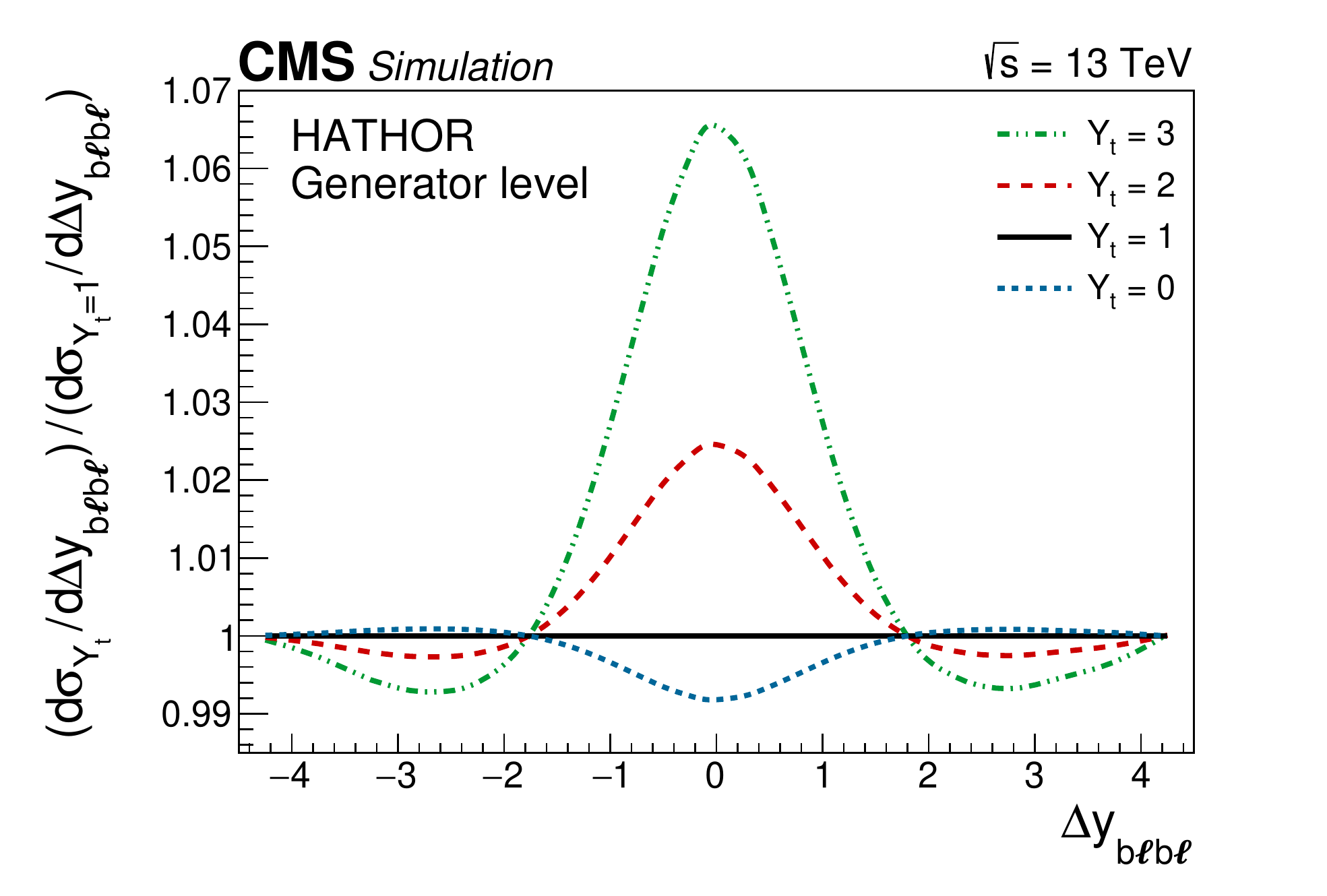}\;
    \includegraphics[width=.48\linewidth]{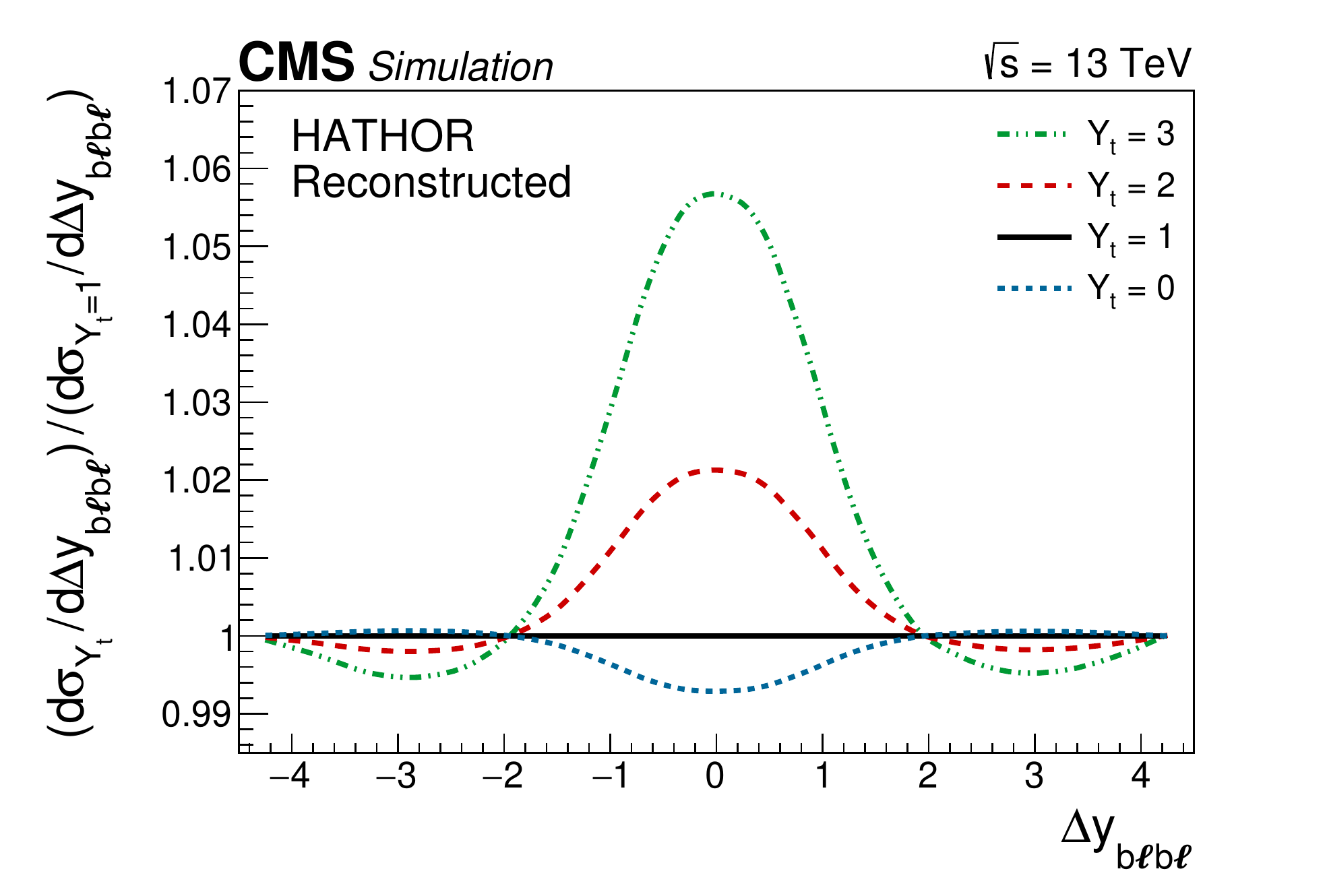}
    \caption{The ratio of kinematic distributions with EW corrections (evaluated for various values of \yt) to the SM kinematic distribution (\yt=1) is shown, demonstrating the sensitivity of these distributions to the Yukawa coupling. The plots on the left show the information at the generator level, while the plots on the right are obtained from reconstructed events. The axis scale is kept the same for the sake of comparison.}
    \label{fig:ratblrec} 
\end{figure*}

The sensitivity of our chosen kinematic variables to $\yt$, before and after reconstruction, is shown in Fig.~\ref{fig:ratblrec}. We see that, in the chosen proxy variables, not much sensitivity is lost in the reconstruction process. This is especially true for the proxy mass observable, \Mbl, providing an advantage over \Mtt, which cannot be reconstructed as accurately.

\subsection{Comparison between data and simulation}
\label{S:control}

\begin{figure*}[tb!]
    \centering
    \includegraphics[width=.47\linewidth]{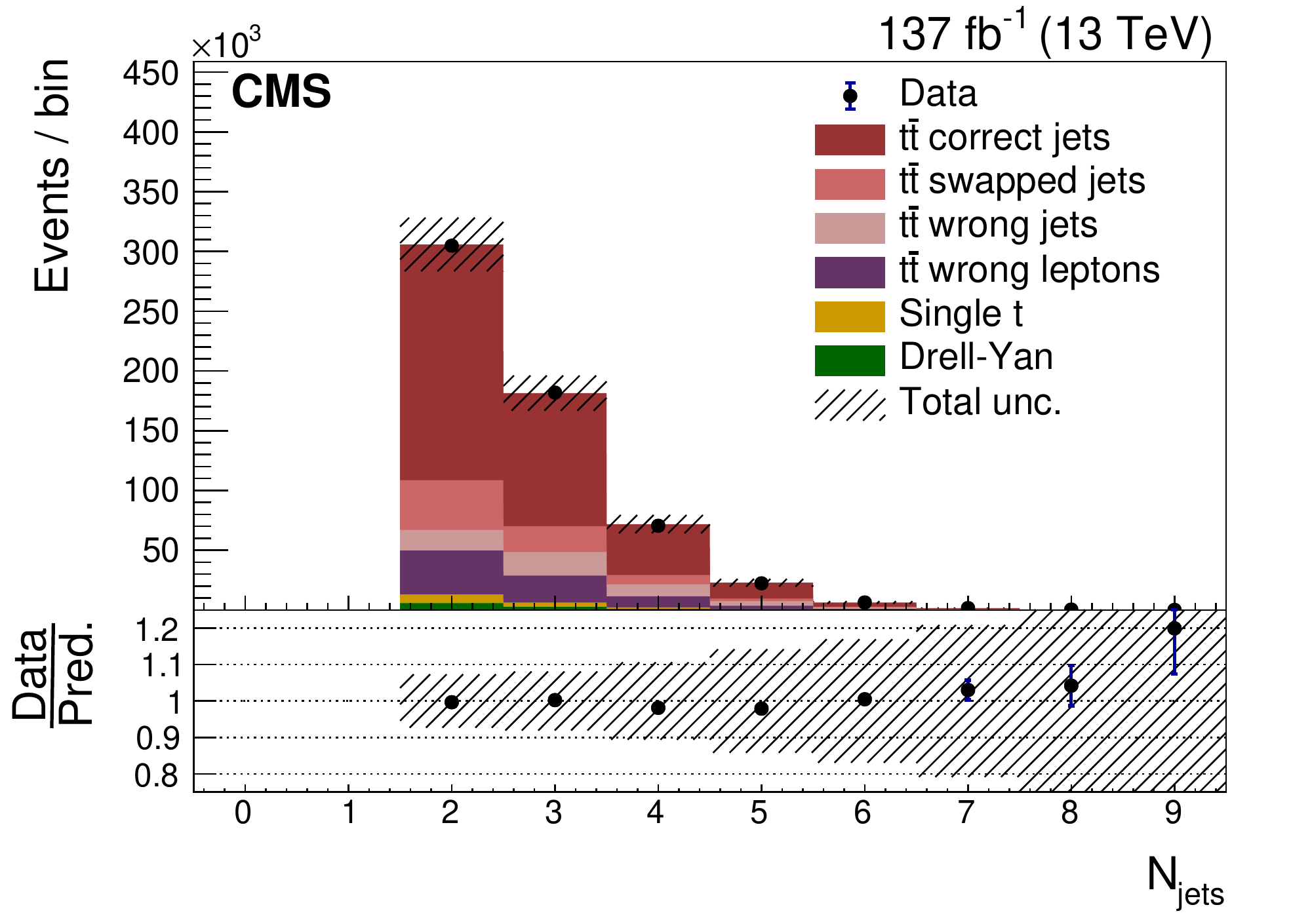}
    \includegraphics[width=.47\linewidth]{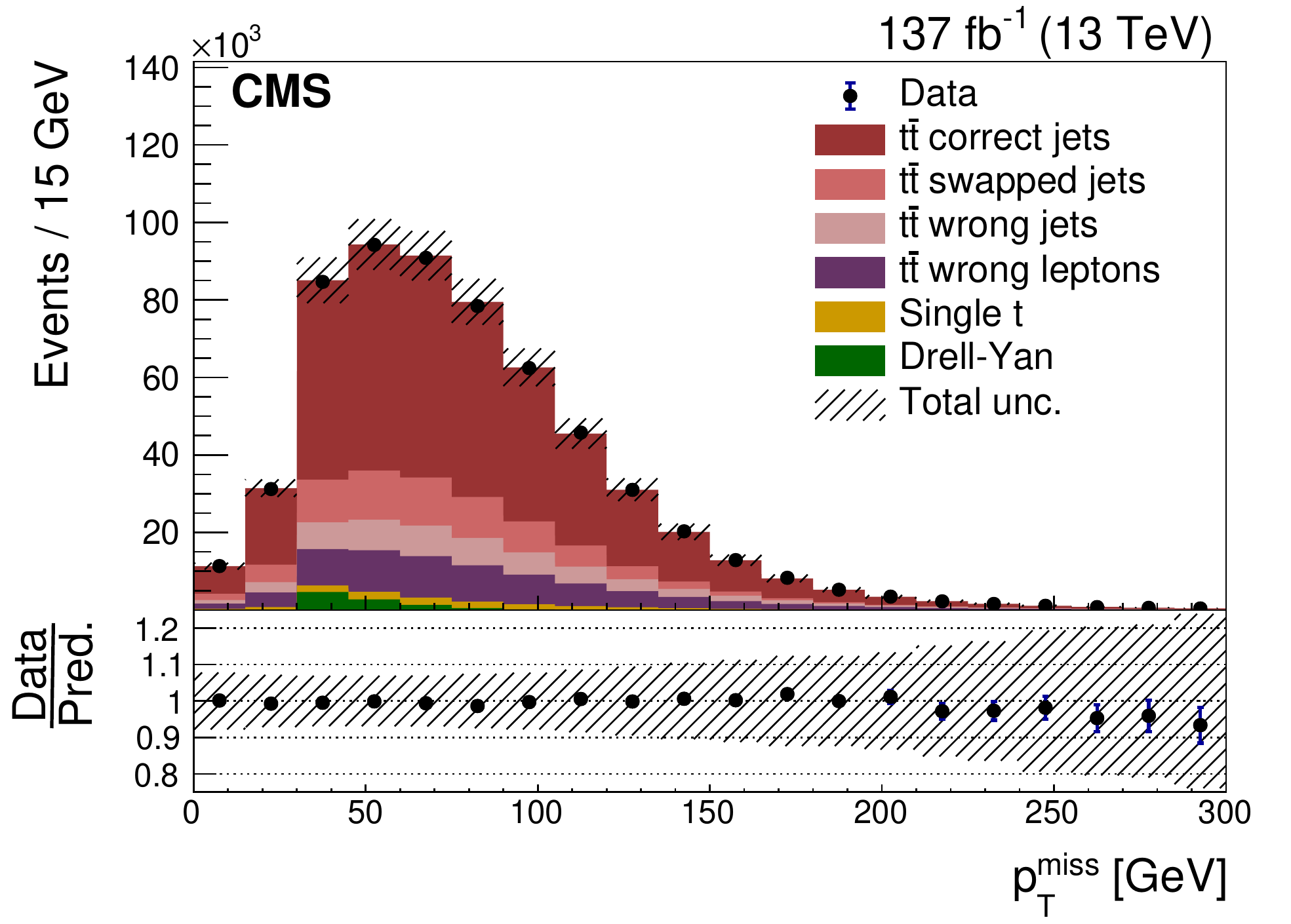}
    \includegraphics[width=.47\linewidth]{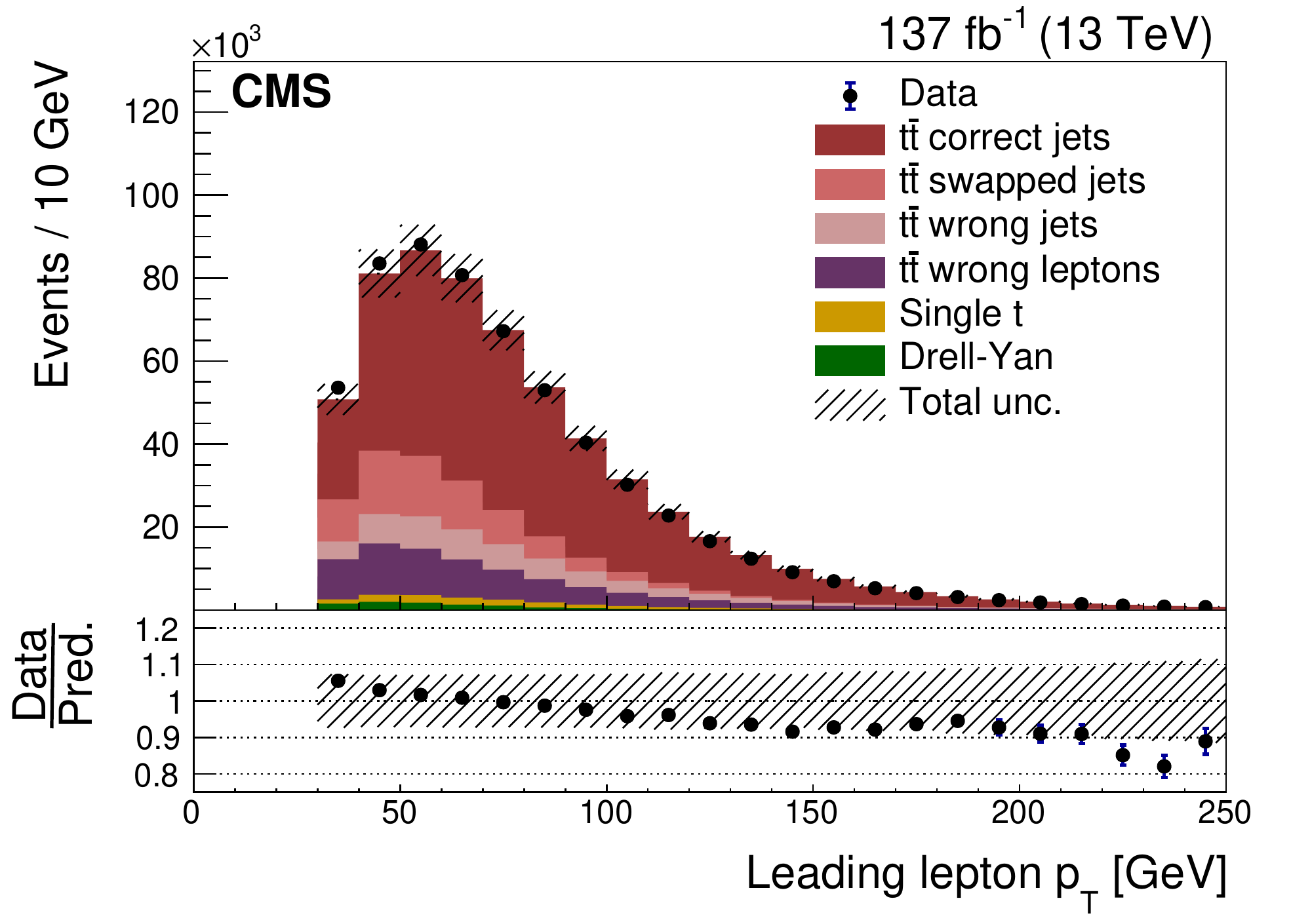}
    \includegraphics[width=.47\linewidth]{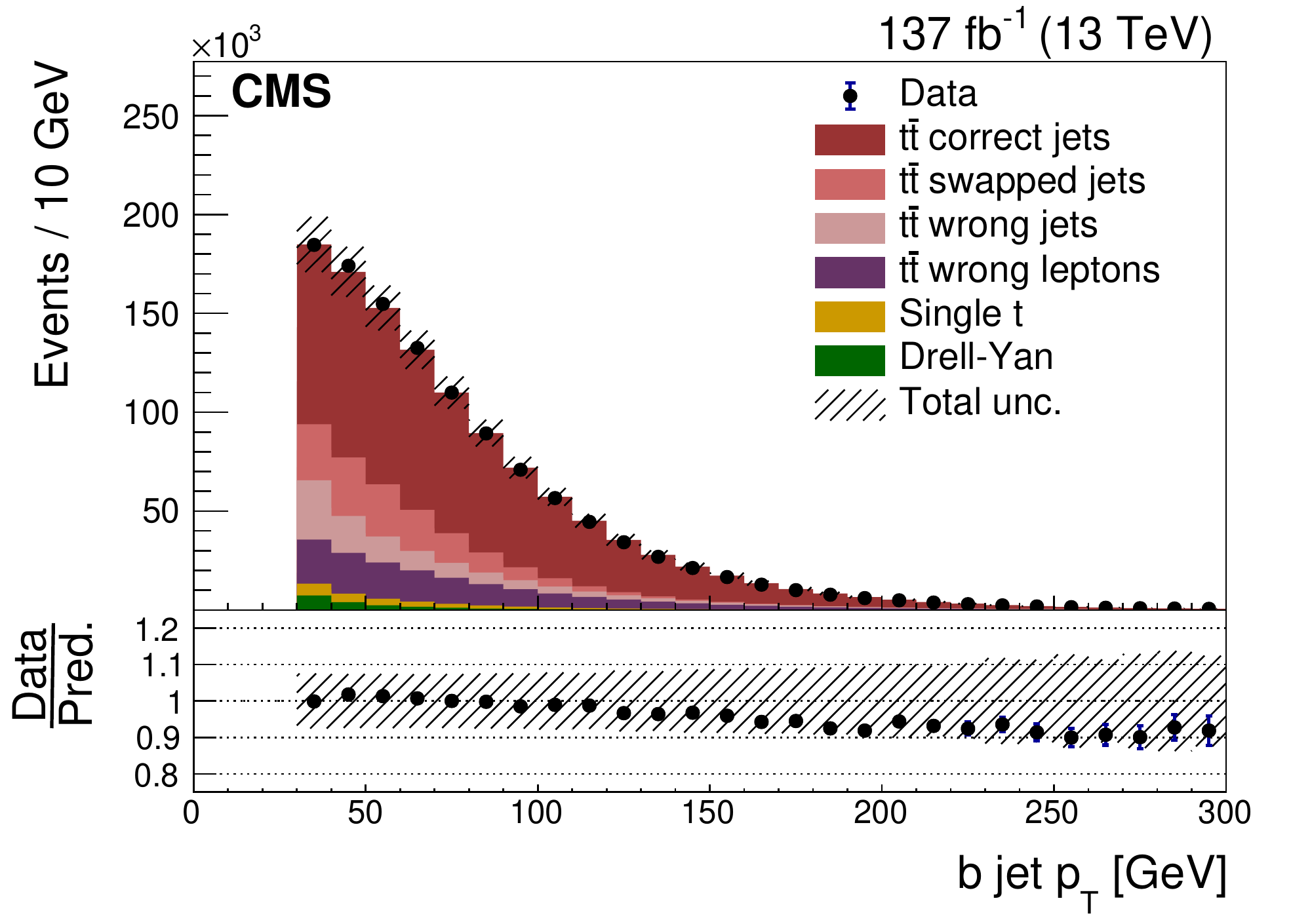}
    \caption{Data-to-simulation comparisons for the jet multiplicity (upper left), \ptmiss (upper right), lepton \pt (lower left), and {\PQb} jet \pt (lower right). The uncertainty bands are derived by varying each uncertainty source up and down by one standard deviation (as described in Section~\ref{S:sys}) and summing the effects in quadrature. The signal simulation is divided into the following categories: events with correctly identified leptons and jets in which jets are correctly assigned (\ttbar correct jets), events with correctly identified leptons and jets in which jets are incorrectly assigned (\ttbar swapped jets), events with correctly identified leptons where the two {\PQb} jets originating from top decays are not identified correctly (\ttbar wrong jets), and lastly events where the identified leptons are not those from \PW boson decay vertices (\ttbar wrong leptons). The lower panels show the ratio of data to the simulated events in each bin, with total uncertainty bands drawn around the nominal expected bin content.}
    \label{fig:dmc}
\end{figure*}

Comparisons between data and simulation are shown in Fig.~\ref{fig:dmc}, where \ttbar events are broken into four categories: events with correctly identified leptons and jets in which jets are correctly assigned (\ttbar correct jets), events with correctly identified leptons and jets in which jets are incorrectly assigned (\ttbar swapped jets), events with correctly identified leptons where the two {\PQb} jets originating from top quark decays are not identified correctly (\ttbar wrong jets), and lastly events where the identified leptons are not those from \PW decay vertices (\ttbar wrong leptons). The majority of events in the last category are \ttbar dilepton decays where a {\PW} boson decay produces a $\tau$ lepton which itself decays leptonically, with a small fraction being misidentified decays in the lepton+jets channel (1\% of the total \ttbar signal). Though all \ttbar events are subject to EW corrections and thus considered as signal, the sensitivity of the reconstructed kinematic variables is generally decreasing among the four categories.

Various observations can be made from Fig.~\ref{fig:dmc}. The agreement between data and simulation appears generally to be within the total uncertainty (discussed further in Section~\ref{S:sys}), and the small overall background rate is apparent. Most events are seen to be associated with zero or one additional jet (beyond the two {\PQb} jets). The effect of the \ptmiss selection requirement can be seen, removing events in the ${\Pe}{\Pe}$ and ${\Pgm}{\Pgm}$ final states in a regime with high Drell--Yan background rates. Single top quark production background rates are seen to vary less steeply as a function of \ptmiss. Looking at the leading lepton \pt, we see that the additional use of a dilepton trigger would not yield a substantial increase in sensitivity. 

A slope is apparent in the ratio of data to the MC prediction in the \pt distributions of leptons and {\PQb} jets. The trends may be related to a previously observed feature of the nominal \POWHEG{}+\PYTHIA simulation, in which a harder top quark \pt distribution is observed than in data (as discussed, \eg, in Ref.~\cite{Otto:17002}). This behavior is the subject of much discussion in the top quark physics community, so we remark on it in this paper despite the fact that we are primarily concerned with other kinematic variables. Fixed-order NNLO calculations are available that generally show a softer top quark \pt spectrum than in the \POWHEG{}+\PYTHIA simulation, which could be seen as evidence that the discrepancy arises from mismodeling in simulation. However, the modeling of \Mtt does not appear to suffer such issues~\cite{Otto:17002}, and we see no evidence that the kinematic variables used in this measurement are not well-described within the included modeling uncertainties. Further discussion of the top quark \pt spectrum in \POWHEG{}+\PYTHIA and its relation to fixed-order NNLO calculations can be found in Section~\ref{S:sys}.

\section{Measurement strategy and statistical methods}
\label{S:stats}

\begin{figure*}[tb!]
    \centering
    \includegraphics[width=.95\linewidth]{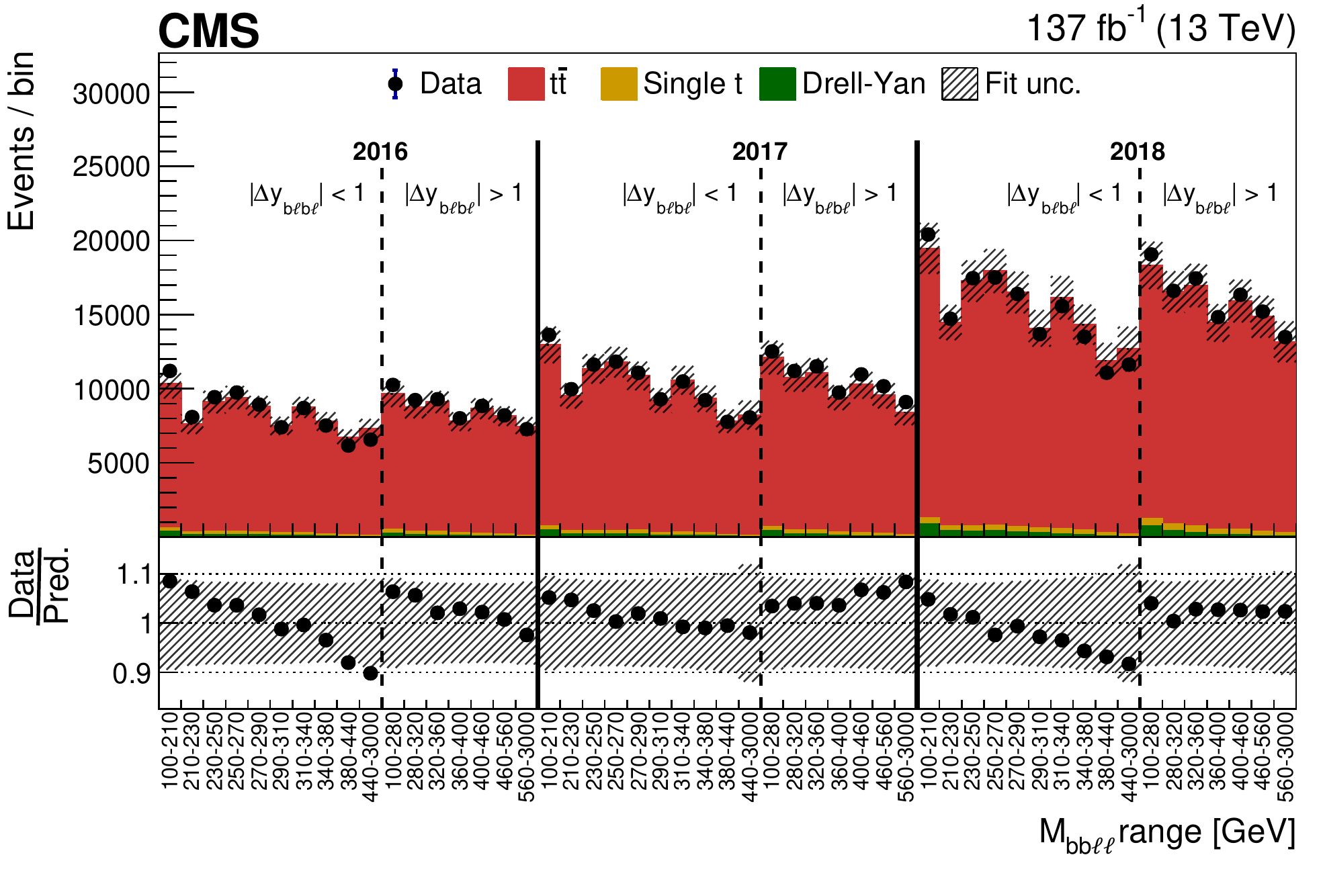}
    \caption{The pre-fit agreement between data and MC simulation in the final kinematic binning. The solid lines divide the three data-taking periods, while the dashed lines divide the two \absDybl bins in each data-taking period, with \Mbl bin ranges displayed on the $x$ axis. The lower panel shows the ratio of data to the simulated events in each bin, with total uncertainty bands drawn around the nominal expected bin content, obtained by summing the contributions of all uncertainty sources in quadrature.}
    \label{fig:ttbinned}
\end{figure*}

After reconstruction, events are binned coarsely in \absDybl and more finely in \Mbl. The binning is chosen to ensure each bin in each data-taking year contains at least on the order of 10\,000 events, as seen in Fig.~\ref{fig:ttbinned}, leading to a low statistical uncertainty and improved uncertainty estimation. 

\begin{figure}
    \centering
    \includegraphics[width=.48\textwidth]{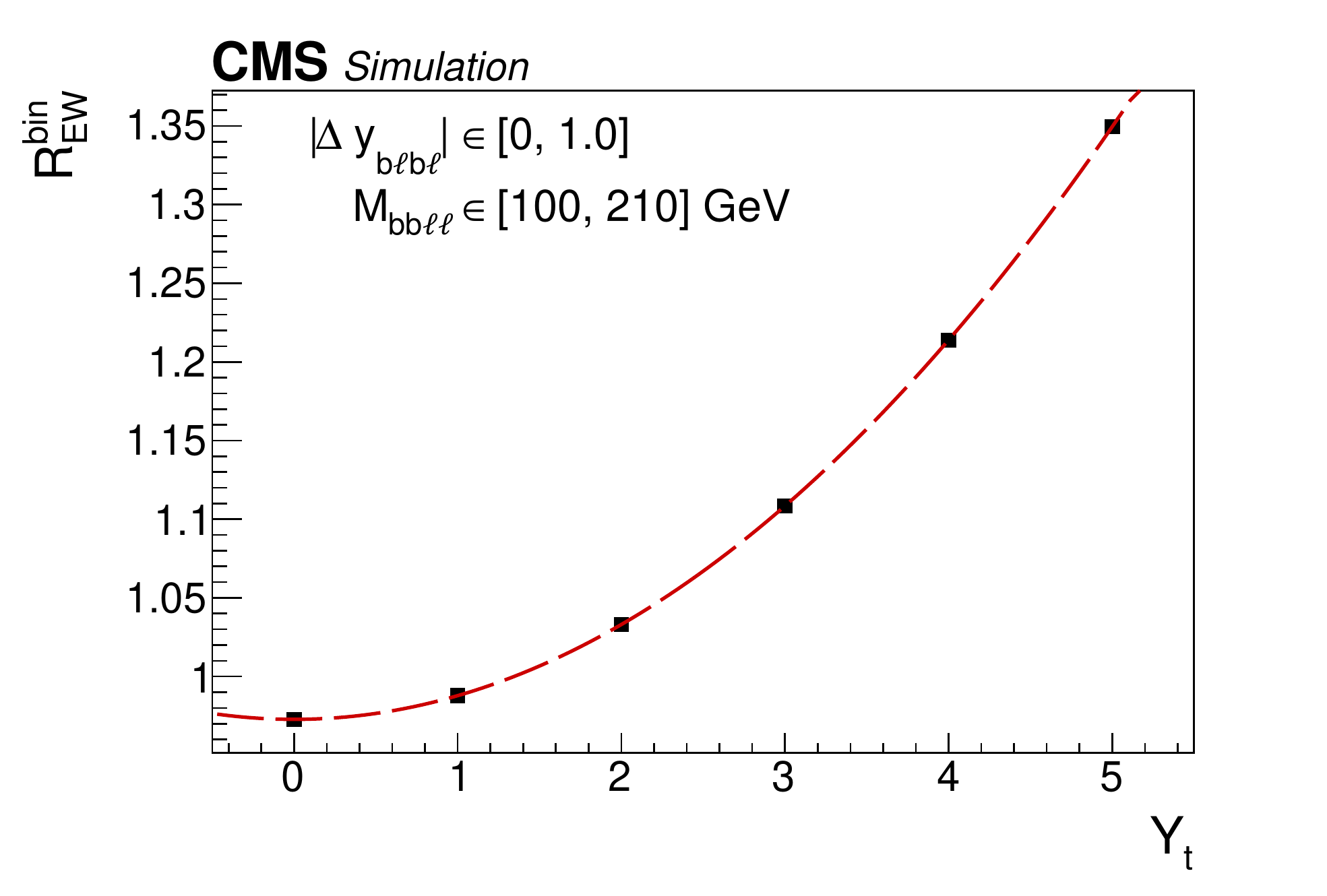}
    \includegraphics[width=.48\textwidth]{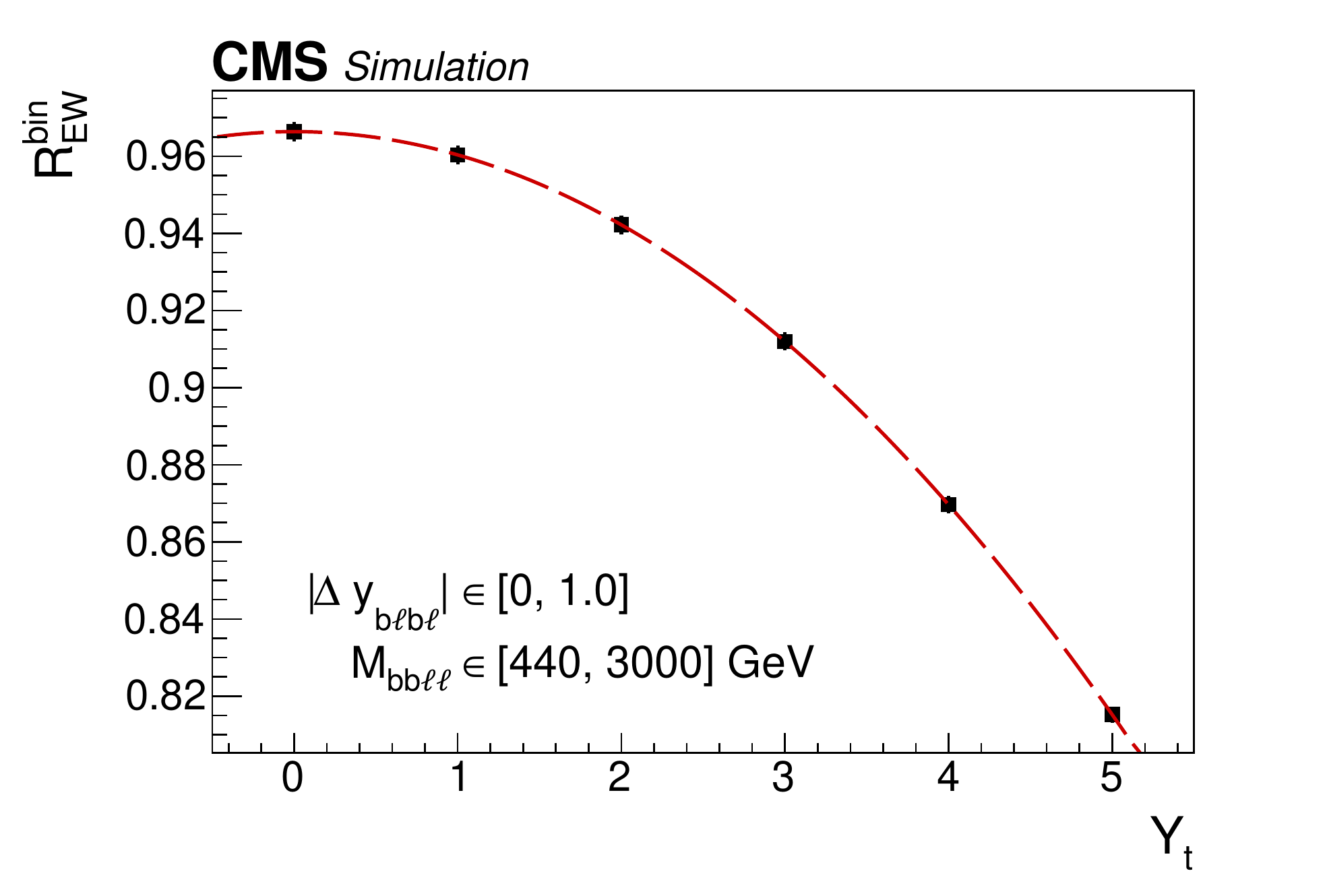}
     \caption{The EW correction rate modifier $R_\mathrm{EW}^\text{bin}$ in two separate ($\Mbl, \Dybl$) bins from simulated 2017 data, demonstrating the quadratic dependence on \yt. All bins have an increasing or decreasing quadratic yield function, with the steepest dependence on \yt found at lower values of \Mbl.}
        \label{fig:quad}
    \end{figure}

In each bin, the expected yield is parametrized as a function of $\yt$. The effect is exactly quadratic, as a consequence of the order at which EW corrections are evaluated. We perform a quadratic fit to extrapolate the effect of the EW corrections on a given bin as a continuous function of $\yt$ (Fig.~\ref{fig:quad}). This correction for each bin can be applied as a rate parameter $R_\mathrm{EW}$ affecting the expected bin content. 

We construct a likelihood function $\mathcal{L}$, 
\begin{equation}
    \mathcal{L} = \left[ \prod_{\text{bin}\in (\Mbl,\absDybl) } \mathcal{L}_\text{bin} \right] \, p(\phi) \, \prod_i p(\theta_i),
\end{equation}
where $\phi$ and $\{\theta_i\}$ are the suite of nuisance parameters associated with individual sources of systematic uncertainty. The distributions $p(\phi)$ and $p(\theta_i)$ are penalty terms which assign {probability} distributions that encode the likelihood the parameters vary from their prior values, as discussed further below. Each bin has an individual Poisson likelihood distribution,
\begin{equation}
    \mathcal{L}_\text{bin} = \mathrm{Poisson}\Big[n^\text{bin}_\text{obs}  \Big\vert s^\text{bin}(\{\theta_i \}) \, R_\mathrm{EW}^\text{bin}(\yt, \phi) +b^\text{bin}(\{\theta_i\}) \Big], 
    \label{eq:likely}
\end{equation}
describing the probability of a bin content to vary from statistical fluctuations. Here $n^\text{bin}_\text{obs}$ is the total observed bin count, with the expected bin count being the sum of the predicted signal yield $s^\text{bin}$ and background yield $b^\text{bin}$. The number of expected signal events is modified by the additional rate parameter $R_\mathrm{EW}$, which depends on the Yukawa coupling ratio $\yt$ and a special nuisance parameter $\phi$ that encodes the uncertainty associated with the multiplicative application of EW corrections derived at order $\mathcal{O}(\alpS^2\alpha)$. The full expression for the rate $R_\mathrm{EW}^\text{bin}$, including this uncertainty term in the bins near the \ttbar production threshold, is given by
\begin{equation}
    R_\mathrm{EW}^\text{bin}(\yt, \phi) = [1+ \delta_\mathrm{EW}^\text{bin}(\yt)]\,[1+\delta_\mathrm{QCD}^\text{bin}\delta_\mathrm{EW}^\text{bin}(\yt)]^\phi,
\end{equation}
where we have defined 
\begin{equation}
 \delta_\mathrm{\mathrm{EW}}^\text{bin}= \frac{n^{\text{bin}}_\mathrm{\textsc{hathor}}-n^\text{bin}_\mathrm{LO\,QCD}}{n^\text{bin}_\mathrm{LO\,QCD}},\  
 \delta_\mathrm{QCD}^\text{bin}= \frac{n^{\text{bin}}_\mathrm{\POWHEG}-n^\text{bin}_\mathrm{LO\,QCD}}{n^\text{bin}_{\POWHEG}}. 
\end{equation}

In the nominal case, we have $R_\mathrm{EW}^\text{bin}(\yt) = 1 +  \delta_\mathrm{EW}(\yt)$. Intuitively, $\delta_\mathrm{EW}$ represents the marginal effect of EW corrections included in \textsc{hathor} relative to the LO QCD calculation, while $\delta_\mathrm{QCD}$ represents the marginal effect of higher-order terms included in the \POWHEG sample relative to the LO QCD calculation. The multiplicative approach to including EW corrections assumes that these two corrections factorize. The quantity $\delta_\mathrm{QCD}^\text{bin}\delta_\mathrm{\mathrm{EW}}^\text{bin}$ represents the cross term arising from the difference in multiplicative and additive approaches. The Gaussian-distributed nuisance parameter $\phi$ modulates the uncertainty generated by this cross term, inducing a bin yield which varies according to a log-normal distribution. We note that the uncertainty in the EW corrections is unique because it depends on the value of \yt at which the EW corrections are evaluated. Thus, it is described by its own term and nuisance parameter $\phi$, separate from other systematic uncertainties. For bins away from the threshold where EW corrections decrease as a function of \yt, we do not include this uncertainty. These bins do not contribute much sensitivity to the measurement, and enter a kinematic regime in which this method of uncertainty estimation is no longer meaningful. At the large values of the Mandelstam variable $s$ that correspond to these bins, the dominant terms contributing to $\delta_\mathrm{EW}$ are Sudakov logarithms resulting from {\PW} and {\PZ} boson exchange. These terms factorize well and do not contribute to the uncertainty we wish to model~\cite{Czakon:2017NLOEW}.

Each nuisance parameter $\theta_j$ corresponding to an overall normalization uncertainty, such as the uncertainty in the integrated luminosity or in cross section values, is assumed to follow a log-normal distribution $p(\theta_j)$. Uncertainties with shape effects associated to nuisance parameters $\{\theta_i\}$ are handled by generating up and down variations of the bin content $s^\text{bin}$ for each $\theta_i$. These variations result from changing the underlying theoretical/experimental sources, which are outlined in Section~\ref{S:sys}, usually by one standard deviation ($\sigma$) based on the uncertainty in our best estimates. These up and down variations are then enforced to correspond to the bin modifiers associated with $\theta_i=\pm1$, while $\theta_i=0$ corresponds to the nominal estimate. The nuisance parameter $\theta_i$ is then taken to follow a Gaussian distribution $p(\theta_i)$ with mean $\mu=0$ and variance $\sigma^2=1$ in the likelihood. The collection of bin modifiers for these up and down variations are referred to as templates, with examples shown in Section~\ref{S:fit}. A vertical template morphing is applied to alter the shape as a function of the underlying nuisance parameter $\theta_i$, where in each bin the modifier is interpolated as a sixth-order polynomial spline for values of $\theta_i\in[-1,1]$ and linearly outside of that region, assuring that $s^\text{bin}(\theta_i)$ remains continuous and twice differentiable.

The measurement of \yt is then performed via a profile likelihood scan, as described in Ref.~\cite{Khachatryan:2014jba__LLscan}. By repeating a maximum likelihood fit over a fine array of fixed values of \yt and comparing to the likelihood at the best fit value, we can use the properties of the maximum likelihood test statistic to evaluate intervals at 68\% and 95\% \CL around the best fit value.

\section{Experimental and theoretical uncertainties}
\label{S:sys}
\subsection{Sources of uncertainty}
The list of uncertainties considered is very similar to that of the previous measurement presented in Ref.~\cite{ytpaper}. The main differences are the lack of QCD multijet background and the use of data from the full Run 2 data-taking period. Full or partial correlations are imposed on the underlying uncertainty sources between data-taking periods where appropriate, as discussed further in Section~\ref{SS:process}. Uncertainties that do not alter the shape of the final distribution are treated as normalization uncertainties, while all others are treated as shape uncertainties on the binned data. Shape effects are considered for the distributions of \ttbar events only, as the contribution of background events is small. Correlations of the uncertainties between different data-taking periods are treated on a case-by-case basis. Because the measurement is more sensitive to shape effects than normalization effects, the uncertainties with the largest magnitude do not necessarily have the largest impact on the measurement sensitivity. By repeating the measurement with any given nuisance parameters frozen at their post-fit values, we are able to evaluate what fraction of the measured uncertainty on \yt is associated to those nuisance parameters.

The dominant experimental uncertainty in this analysis comes from the calibration of the detector jet energy response. Corrections to the reconstructed jet energies are applied as a function of \pt and $ \eta$. We follow the standard approach outlined in Ref.~\cite{JES} to consider 26 separate uncertainties that are typically involved in determining these calibrations. In this approach, the uncertainty in the resolution of the jet reconstruction is also considered in addition to the energy response. The effect of these uncertainties is propagated to the reconstruction of \ptmiss. These effects account for approximately $7\%$ of the total uncertainty on \yt in the final measurement.

Other experimental sources of uncertainty are comparatively minor. The overall uncertainty in the integrated luminosity of 2.5, 2.3, and 2.5\% is included as a normalization uncertainty applied to all signal and background events in 2016, 2017, and 2018, respectively~\cite{lumi16,lumi17,lumi18}. The uncertainty in the number of pileup events included in simulation is assessed by varying the inelastic cross section, 69.2\unit{mb}, by 4.6\%~\cite{InelasticXS}.

Efficiencies in {\PQb} jet identification and misidentification are corrected to match data~\cite{btagRun2}. While this source is treated as a shape effect, the uncertainty manifests approximately as an overall normalization effect on the signal of around 3\%, and contributes only about 1\% of the final uncertainty on \yt.

Similarly, scale factors are applied in bins of \pt and $\eta$ to correct simulated efficiencies of lepton reconstruction, identification, isolation, and triggers to match data. These are derived from a fit using the tag-and-probe method using {\PZ} boson decays~\cite{Khachatryan:2010xn__DYtnp,electronPerformance8TeV,muonPerformance13TeV}. This fit accounts for the uncertainty from the limited number of events in the data sample as well as differences in performance based on the jet multiplicity. Overall, the effect is assessed to be below 2\%.

As a standard technique to estimate the contributions of higher-order QCD terms at the ME level, the renormalization scale \mur and factorization scale \muf are each varied up and down in the \POWHEG simulation by a factor of 2. Templates are generated for the individual variation of \mur and \muf, as well as an additional template for the simultaneous variation of the two scales together (up and down), leading to three separate shape uncertainties in total. Since an NNLO \ttbar cross section is already used to improve the normalization of the MC simulation, the normalization effect induced by the scale variations is overestimated. As we include a separate uncertainty on the cross section normalization, the overall normalization effect is therefore removed entirely from the scale variation templates, which are normalized to the nominal sample. The resulting shape effects remain significant and these are among the limiting uncertainties in the fit, contributing about 7\% of the total measurement uncertainty.

A 5\% normalization uncertainty is assumed in the \ttbar cross section, which covers expected contributions from the higher-order terms not included in the NNLO+NNLL cross section calculation~\cite{Czakon:2011xx}, giving a more realistic normalization uncertainty than the variation of \mur and \muf in \POWHEG. The backgrounds in this analysis are small enough ($\approx$2\% sample composition each) that we do not generate templates for their response to individual systematic uncertainties. A 15\% normalization uncertainty is included on single top quark MC samples, which covers the expected ME scale variation and the jet energy correction uncertainties associated with these samples. The Drell--Yan and diboson MC samples are assigned a 30\% normalization uncertainty, to cover the larger ME scale variation uncertainties associated with these LO simulations. The background normalizations can alter slightly the expected shape of the data, but are not among the most impactful uncertainties.

We include an uncertainty in the EW corrections, based on our methods for generating and applying these additional terms, as outlined in Sections~\ref{S:simulation} and~\ref{S:stats}. Like the scale variations, this uncertainty is designed to cover higher-order effects at the ME level, specifically those arising from diagrams of order $\alpS^3 \alpha$. It places an uncertainty on $R_\mathrm{EW}^\text{bin}$ of 10--40\% in the applicable bins, which translates to a small overall uncertainty in bin rate unless the corrections are evaluated at a value of \yt far from the SM expectation. This helps ensure that we do not fit an artificially high value of \yt by ignoring higher-order diagrams. This represents one of the most significant uncertainties in the fit, accounting for approximately $8\%$ of the final measurement uncertainty. It is also observed to primarily affect the lower bound of the measurement, thus reducing our ability to distinguish between values of $\yt<1$. 

The uncertainty in modeling the initial- and final-state radiation in the parton shower algorithm is assessed by varying the value of the renormalization scales in the initial- and final-state radiation by a factor of two. These are among the most limiting modeling uncertainties in the measurement, contributing about $8\%$ of the total measurement uncertainty. Uncertainties for other parameters in the parton shower description are considered separately. The $h_\mathrm{damp}$ parameter, which controls the ME to parton shower matching in \POWHEG{}+\PYTHIA, is set to the nominal value of $h_\mathrm{damp}= 1.58 \, m_{\PQt}$ (1.39 $m_{\PQt}$) in 2016 (2017--18). Dedicated MC samples are generated with this parameter varied down to $1  \,m_{\PQt}$ ($0.874 \,m_{\PQt}$) and up to $2.24 \,m_{\PQt}$ ($2.305\, m_{\PQt}$) in 2016 (2017--18), in order to estimate the effect of this uncertainty. Dedicated MC samples are also generated with variations of the \PYTHIA underlying-event tune. The uncertainties due to the choice of $h_\mathrm{damp}$ and the underlying-event tune are very minor compared to the parton shower scale variations.

Dedicated MC samples are generated with the top quark mass varied up and down by 1\GeV from the nominal value $m_{\PQt}=172.5\GeV$ to estimate the effect of the uncertainty in the measured mass value. While this uncertainty has a significant shape effect, it ultimately accounts for only about $1\%$ of the total measurement uncertainty. It should be noted that, although the mass and Yukawa coupling are generally treated as independent in this measurement, varying the mass will slightly modify the definition of $\yt=1$. However, this effect, which is below 1\%, is much smaller than the sensitivity of the measurement and can therefore be ignored. 

The NNPDF sets~\cite{NNPDF} contain 100 individual variations as uncertainties. Following the approach in Ref.~\cite{ytpaper}, similar variations are combined to reduce the number of variations to a more manageable set of 10 templates.  The variation of the strong coupling $\alpS$ used by NNPDF is treated separately from the other PDF variations. The effect of uncertainties in the PDF set is typically smaller than 1\%, and together they account for roughly 2\% of the total measurement uncertainty.

The branching fraction of semileptonic {\PQb} hadron decays affects the {\PQb} jet response. The effect of varying this quantity within its measured precision~\cite{PDG} is included as an uncertainty, which has a small effect relative to other modeling uncertainties. 

The momentum transfer from {\PQb} quarks to {\PQb} hadrons is modeled with a transfer function dependent on $x_{\PQb}=\pt(\PQb\text{-hadron})/\pt(\PQb\text{-jet})$. To estimate the uncertainty, the transfer function is varied up and down within uncertainty of the Bowler--Lund parameter~\cite{Bowler} in \PYTHIA. The resulting effect is included by modifying event weights to reproduce the appropriate transfer function. This has a noticeable shape effect of the order~4\%, but was not found to be a leading uncertainty in the fit.

In some measurements performed strictly in the context of the SM (for example, in Ref.~\cite{topPTexample}), an additional uncertainty is included to account for an observed difference in the top quark \pt distribution between data and \POWHEG{}+\PYTHIA simulation. As the measurement presented here is sensitive to anomalously high values of \yt, we do not want to include any additional uncertainties which explicitly enforce agreement between SM simulation and the data, as this could reduce our sensitivity to deviations from the SM.

With this in mind, studies were performed comparing different simulations to assess whether top quark \pt modelling disagreements necessitated the inclusion of any additional  uncertainties. Fixed-order calculations were studied for \ttbar production at NNLO, which generally show better agreement with the top quark \pt spectrum observed in data (see, for example, Refs.~\cite{Grazzini:2017mhc__matrix,Otto:17002}). Specifically, differential cross sections in top quark \pt and \Mtt were studied using publicly available \textsc{fastNLO} tables~\cite{Czakon:fnlo1,Czakon:fnlo2}, as well as multidifferential cross sections~\cite{Czakon:2Dfnlo}. Such NNLO calculations use a different choice of dynamical scale in evaluating the top quark \pt versus other kinematic variables, lending them an edge in precision over full event simulation. We find that the predictions from \POWHEG{}+\PYTHIA samples are consistent with the differential and multidifferential cross sections from Refs.~\cite{Czakon:fnlo1,Czakon:fnlo2,Czakon:2Dfnlo} involving \Mtt and \Dytt , within modeling uncertainties. These distributions appear consistent with the data as well, as previously observed in Ref.~\cite{Otto:17002}. By comparison, the top quark \pt distribution evaluated at NNLO from Refs.~\cite{Czakon:fnlo1,Czakon:fnlo2} shows more substantial disagreement with \POWHEG{}+\PYTHIA simulations. We conclude that the variables relevant to our measurement technique appear sufficiently well described by \POWHEG{}+\PYTHIA simulations, and differences with relevant NNLO calculations should be covered by the standard uncertainty estimation techniques. However, analyses that are more specifically sensitive to the top quark \pt distribution should take care in addressing this discrepancy when using \POWHEG{}+\PYTHIA samples.

Information about the magnitudes and effects of significant uncertainties can be found in Table~\ref{tab:uncs}.

\begin{table*}[htb]
\centering
\topcaption{The effect of all significant normalization (norm.) and shape uncertainties is summarized. Uncertainties are grouped into categories based on their physical origin, and the approximate effect on sample yield is stated. Additionally, the fit is repeated with each category frozen to their post-fit values, in order to assess the reduction of total fit uncertainty resulting from their removal (rightmost column). Minor uncertainties with $<1\%$ effect on sample yield are excluded from this summary.}
\newcolumntype{x}{D{,}{\,\%\,}{6.3}}
\begin{scotch}{l c x x}
Uncertainty category & \multicolumn{1}{c}{type} & \multicolumn{1}{c}{effect on yield} & \multicolumn{1}{c}{reduction in fit unc.} \\
\hline 
\rule{0pt}{1.005\normalbaselineskip}\ttbar cross section 			& norm. & \textrm{5}, 			& <1, \\
background norm.				& norm. & \textrm{0--1.5},		& \approx 1, \\
luminosity 						& norm. & \textrm{2.3--2.5},	& <1, \\
jet energy corrections 			& shape & \textrm{0--4}, 		& 7.4, \\
EW correction unc. ($\phi$) 	& shape & \textrm{(0--40},)\,\delta_\mathrm{EW} & 7.6, \\
ME scales 						& shape & \textrm{0--5},		& 7.3, \\
parton shower scales 			& shape & \textrm{0--4}, 		& 7.7, \\
NNPDF uncertainties 			& shape & \textrm{0--3}, 		& 1.9, \\
top quark mass 					& shape & \textrm{0--2.5},	& 1.3, \\
\cPqb tagging efficiency	 	& shape & \textrm{2--2.5}, 	& \approx 1, \\
\cPqb mistagging efficiency	 	& shape & \textrm{0--0.5},	& <1, \\
lepton scale factors 			& shape & \textrm{0--2}, 		& \approx 1, \\
\cPqb fragmentation 			& shape & \textrm{0--5}, 		& <1, \\
\cPqb hadron branching frac.	& shape & \textrm{1--2}, 		& <1, \\
pileup 							& shape & \textrm{0--0.5}, 	& <1, \\
\end{scotch}
 \label{tab:uncs}
\end{table*} 

\subsection{Treatment of systematic uncertainties}
\label{SS:process}

\begin{figure}[hbt]
    \centering
    \includegraphics[width=\cmsFigWidth]{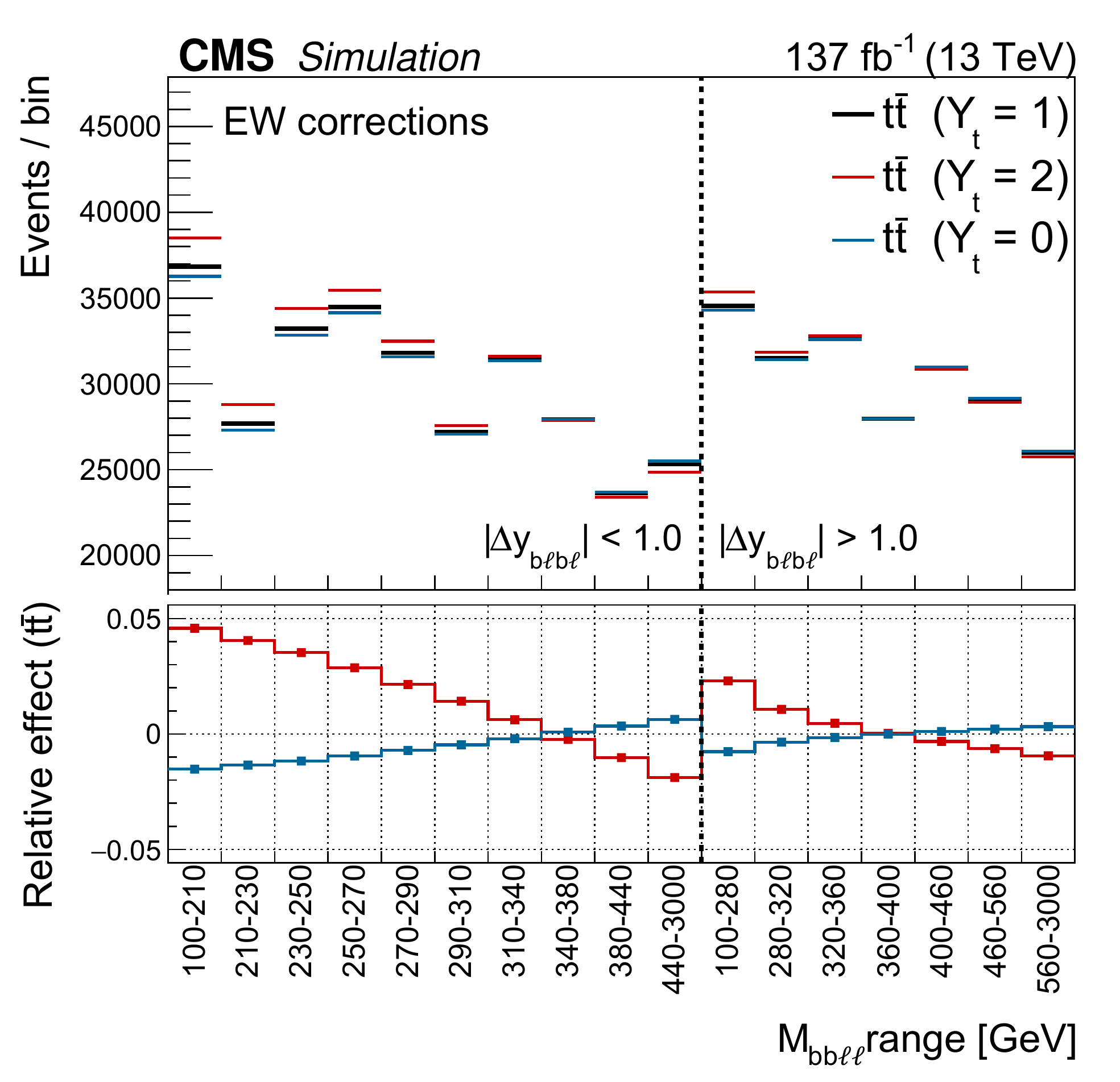} 
    \caption{The effect of the Yukawa parameter \yt on reconstructed event yield in the final binned distributions. The variation of \yt induces a shape distortion in the kinematic distributions. The marginal effect relative to the standard model expectation \yt=1 is visualized in the lower panel.}
    \label{fig:yuktemplate}
\end{figure}

In this analysis, the effect of the parameter of interest \yt manifests itself as a smooth shape distortion of the kinematic distributions, as shown in Fig.~\ref{fig:yuktemplate}. Although the nuisance {parameters} describing the sources of uncertainty should induce smooth shape effects as well, their effects are sometimes obscured by statistical noise or imprecise methods of estimation. This is noticeable for the uncertainties associated with the jet energy scale, jet energy resolution, parton shower modeling, pileup reweighting, and top quark mass. For these templates only, we apply a one-iteration LOWESS algorithm~\cite{LOWESS} to smooth the templates and remove fluctuations that may disturb the fit. The underlying-event tune and $h_\mathrm{damp}$ uncertainties in the parton showering are small enough for their shapes to disappear into statistical noise, and are therefore treated only as normalization uncertainties. 

Most templates are also symmetrized, by taking the larger effect of the up and down variations in each bin and using this magnitude for both. This step helps ensure a stable minimum in the likelihood fit, but is skipped for the templates whose natural shape effect is notably asymmetric. In the few cases where this may be an overly conservative approach, it nonetheless guarantees the performance and reliability of the minimization procedure, and has little effect on the final result. 

Full or partial correlations between the 2016, 2017, and 2018 data analyses are assumed for many uncertainties. In general, the theoretically motivated uncertainties are considered fully correlated between years. Exceptions are made in cases where modeling differed between years. The PDF uncertainties cannot be correlated between 2016 and other data-taking periods, as the PDF sets used for simulation were changed to a newer version. Due to changes in the \PYTHIA tune following 2016, the nominal scales used initial-state radiation and final-state radiation differ after 2016, so those uncertainties are treated as only partially correlated between 2016 and other data-taking periods. The modeling of these uncertainties differs in the 2016 simulation, so the associated nuisance parameter in this year is either partially or fully decorrelated from those in the other years. Additionally, uncertainties whose effects disappear into statistical noise due to limited MC sample size (underlying-event tune and $h_\mathrm{damp}$) are converted to uncorrelated normalization uncertainties. 

Some experimental uncertainties can be broken into components, which are either fully correlated or uncorrelated between years (large jet energy scale contributions and integrated luminosity). The uncertainty in the number of pileup events is considered fully correlated as it is evaluated by varying the total inelastic cross section. For minor uncertainties from jet and lepton scale factors, which have both correlated and statistical components, a 50\% correlation is assumed between years. Lastly, the jet energy resolution uncertainties are treated as uncorrelated between years. 

\section{Results}

\label{S:fit}
\begin{figure}[ht!]
    \centering
    \includegraphics[width=\cmsFigWidth]{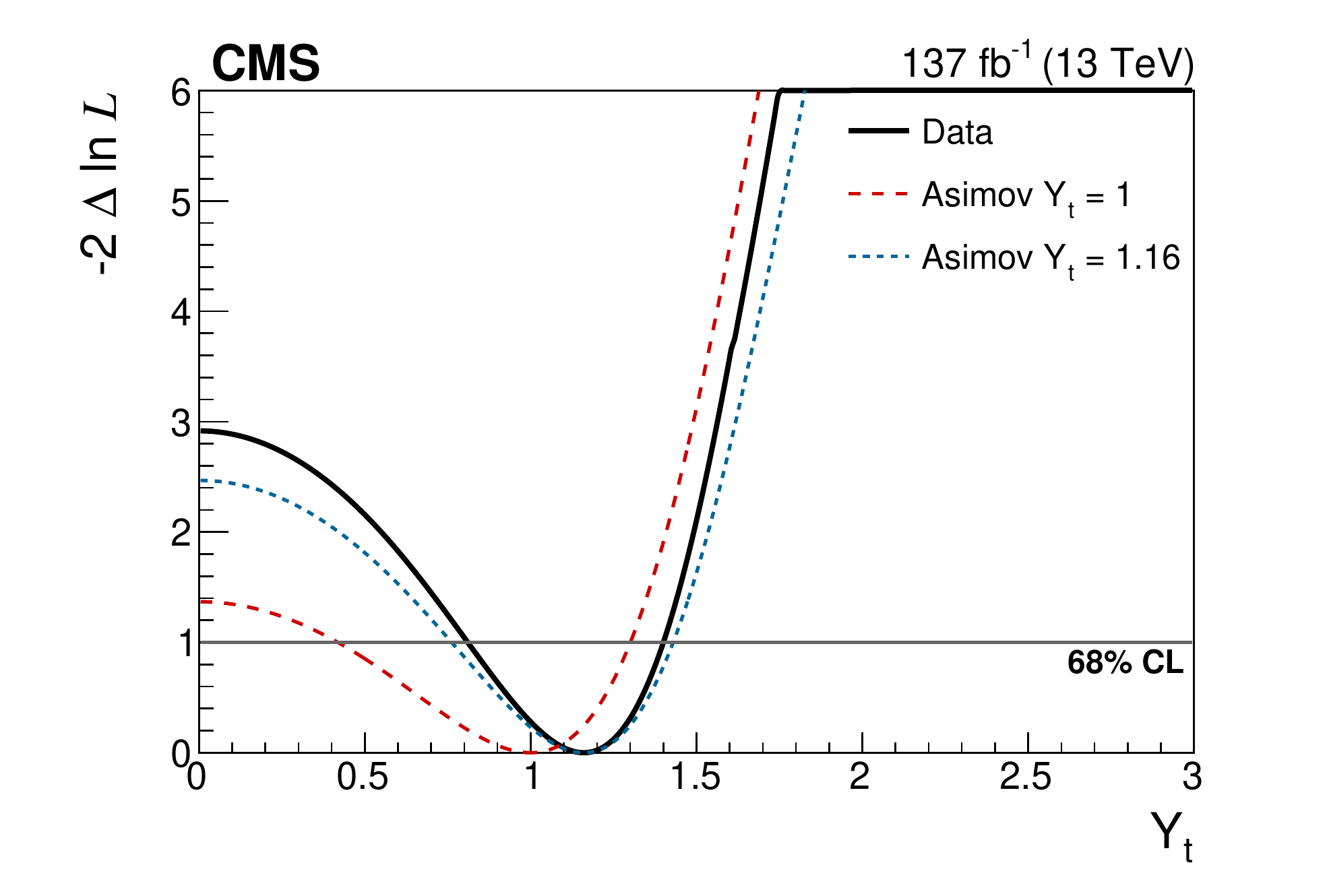}
    \caption{The result of a profile likelihood scan, performed by fixing the value of $\yt$ at values over the interval $[0, 3]$ and taking the ratio of $-2\ln(\mathcal{L}(\yt))$ to the best fit value $-2\ln(\mathcal{L}(\hat{Y}_{\PQt}))$. The expected curves from fits to simulated Asimov data are shown produced for the SM value $\yt=1.0$ (dashed) and for the final best fit value of $\ytbestfit$ (dotted).} 
    \label{fig:scan}
\end{figure}

\begin{figure*}[h!]
    \centering
    \includegraphics[width=0.95\linewidth]{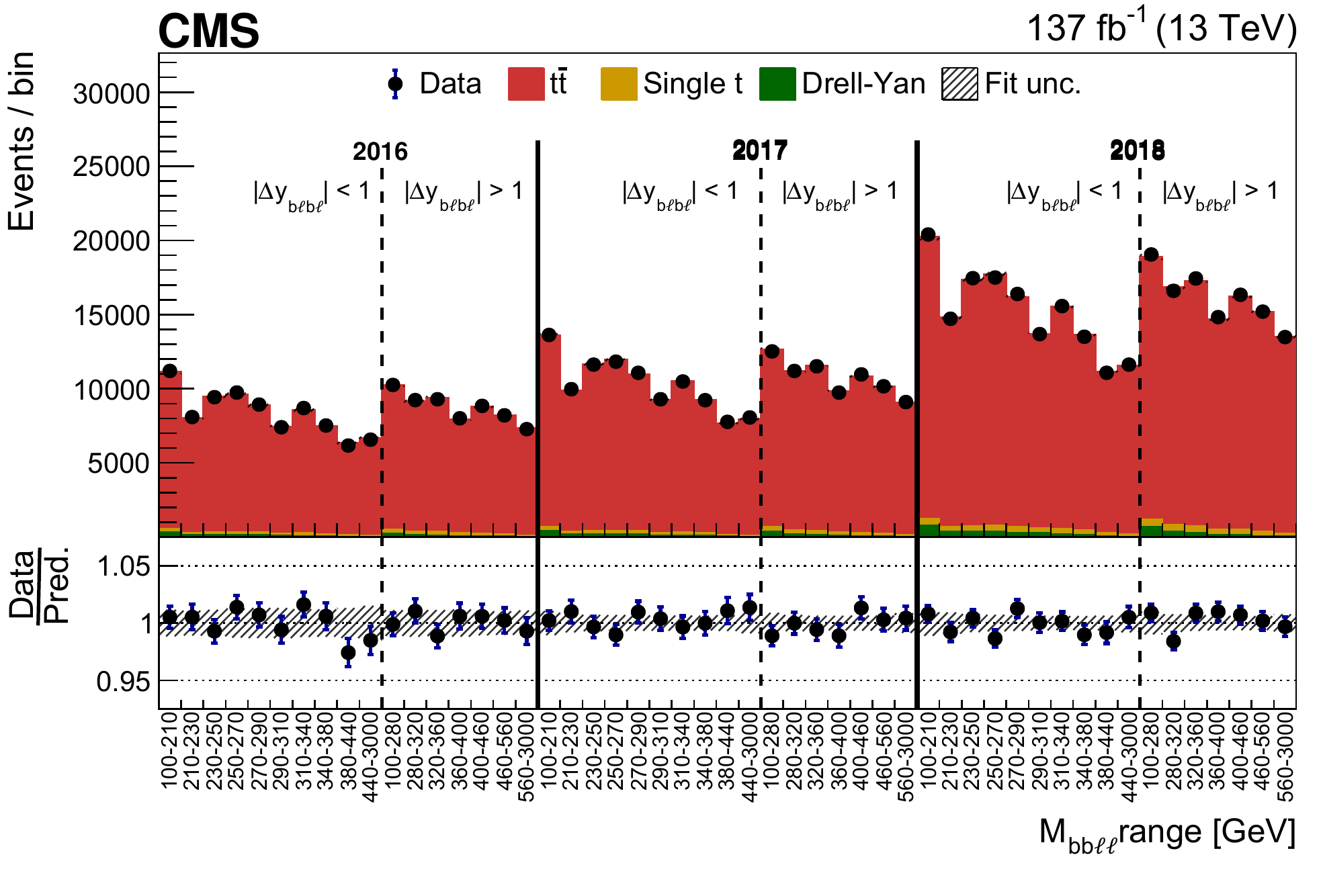}
    \caption{The comparison between data and MC simulation at the best fit value of $\ytbestfit$ after performing the likelihood maximization, with shaded bands displaying the post-fit uncertainty. The solid lines separate the three data-taking periods, while the dashed lines indicate the boundaries of the two \absDybl bins in each data-taking period, with \Mbl bin ranges displayed on the $x$ axis. The lower panel shows the ratio of data to the simulated events in each bin, with total post-fit uncertainty bands drawn around the nominal expected bin content.} 
    \label{fig:postfit}
\end{figure*}

\begin{figure*}[tb!] \centering

\includegraphics[width=.49\linewidth]{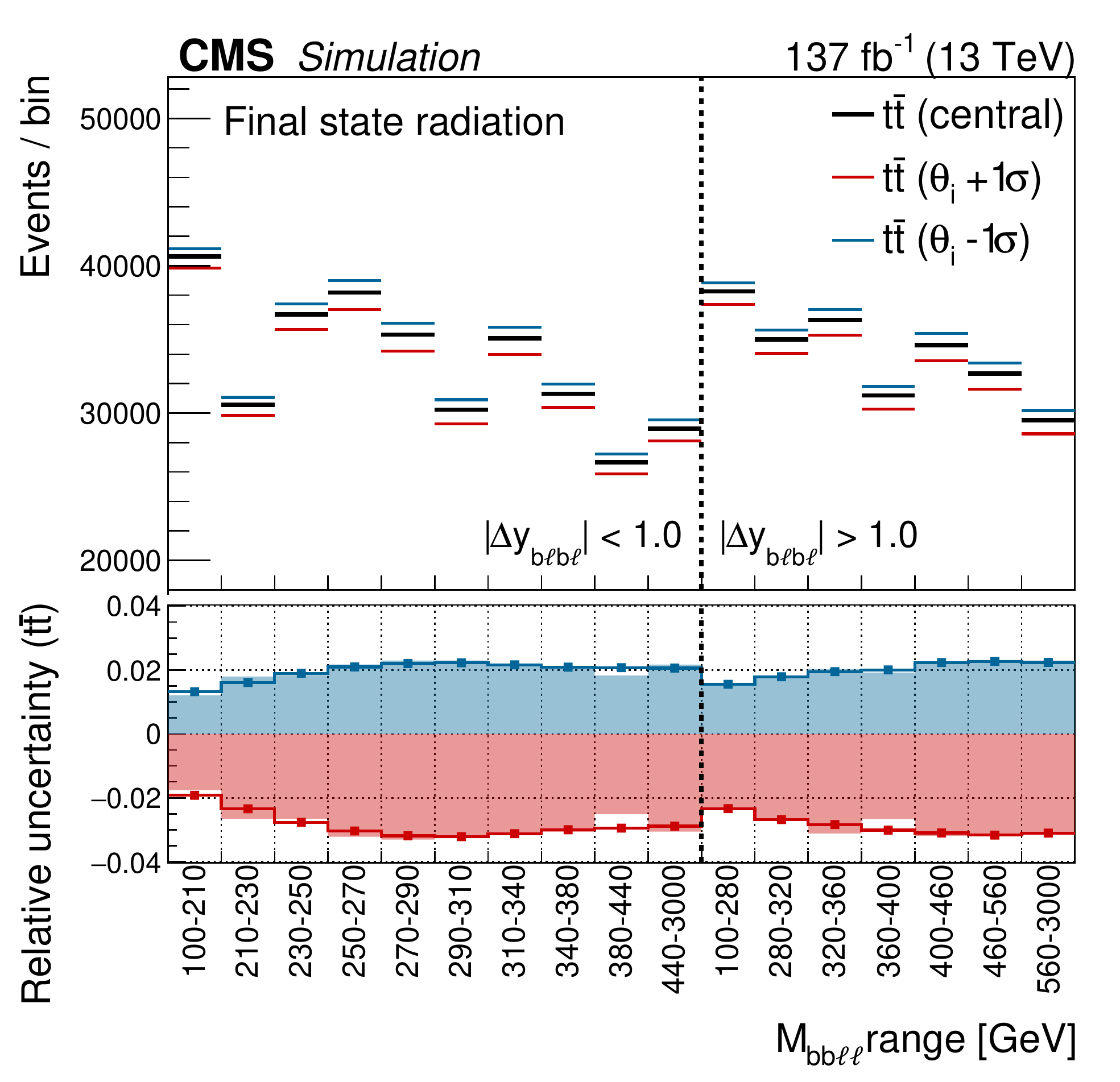}
\includegraphics[width=.49\linewidth]{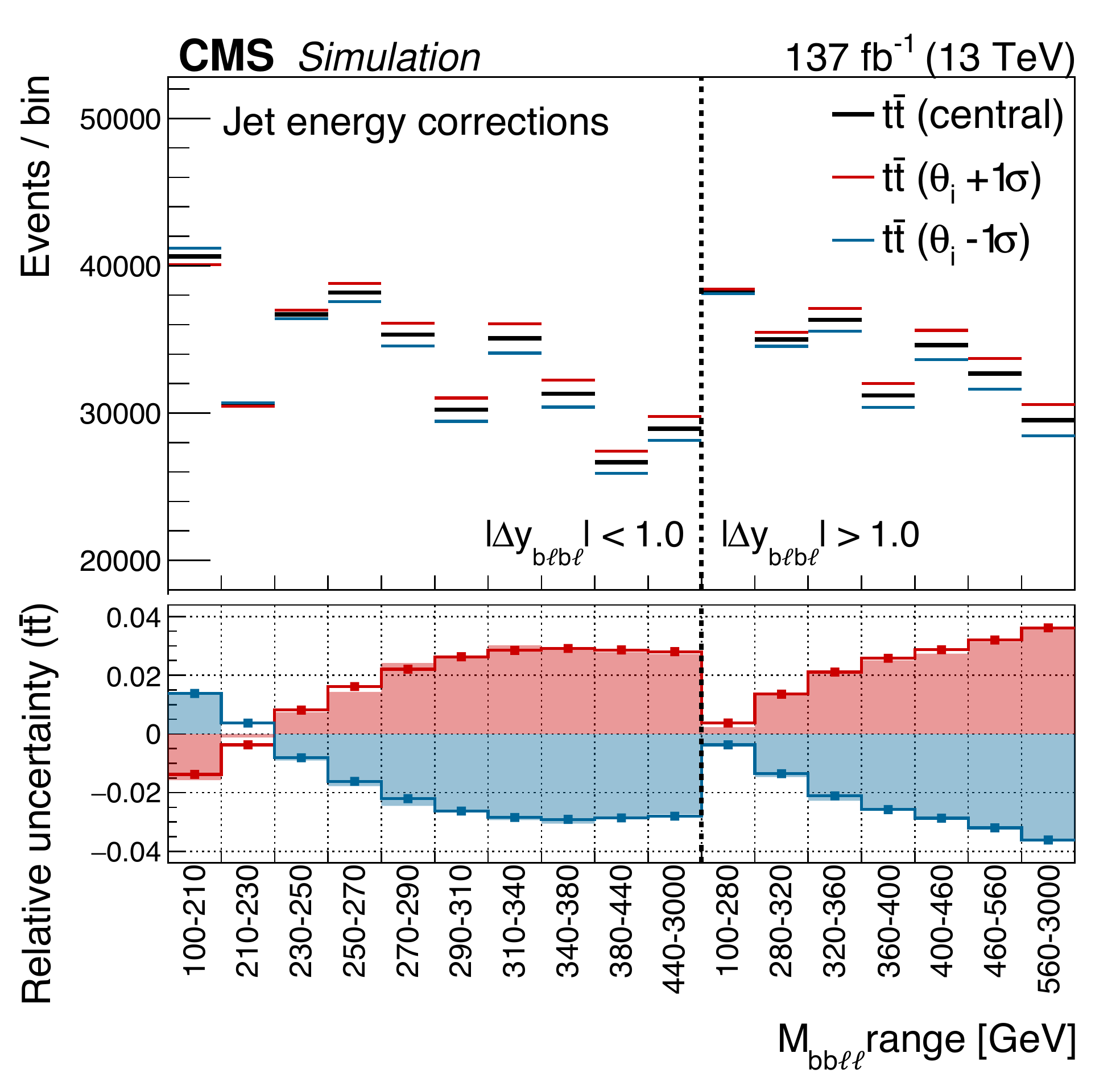}

\includegraphics[width=.49\linewidth]{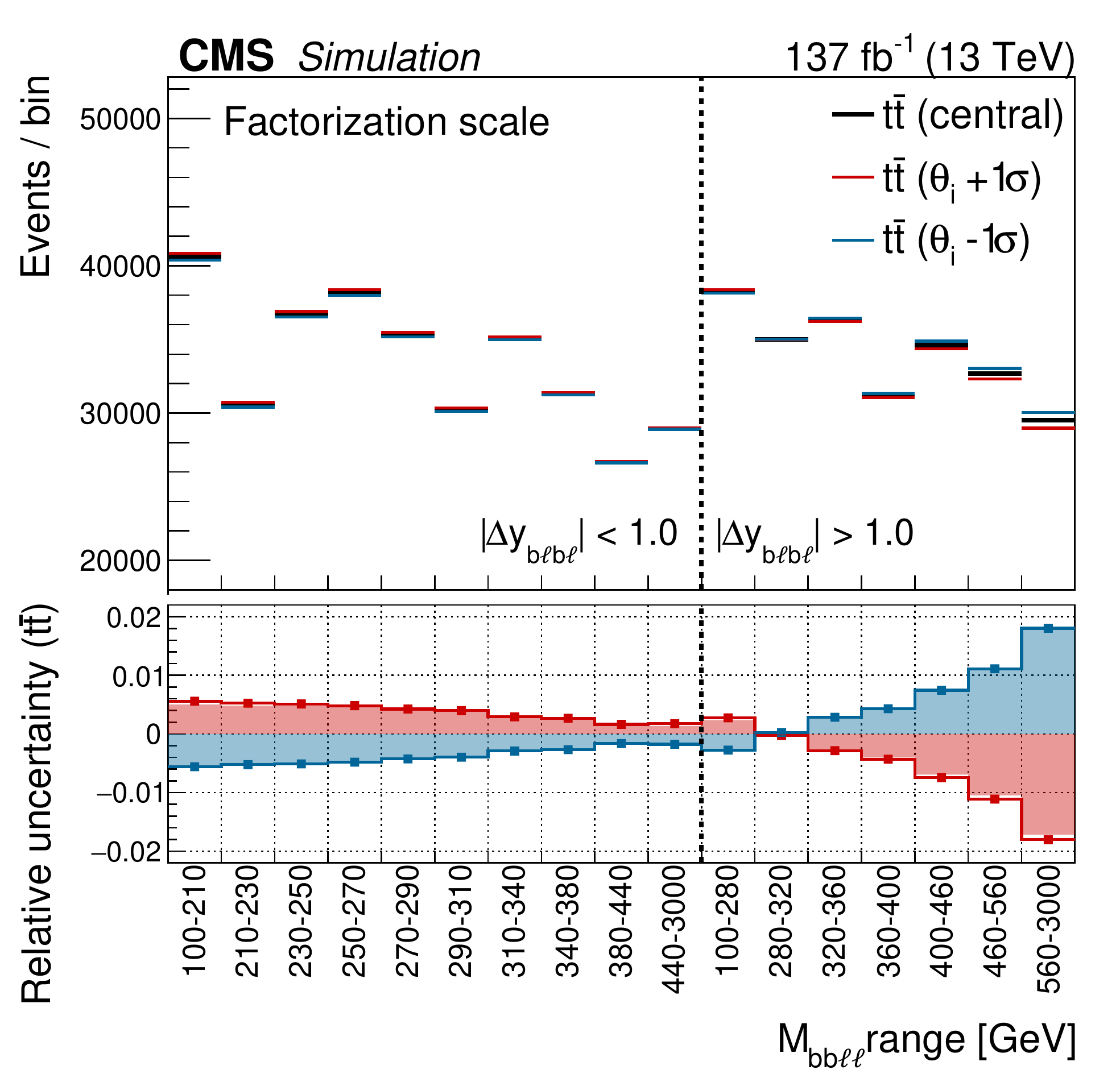}
\includegraphics[width=.49\linewidth]{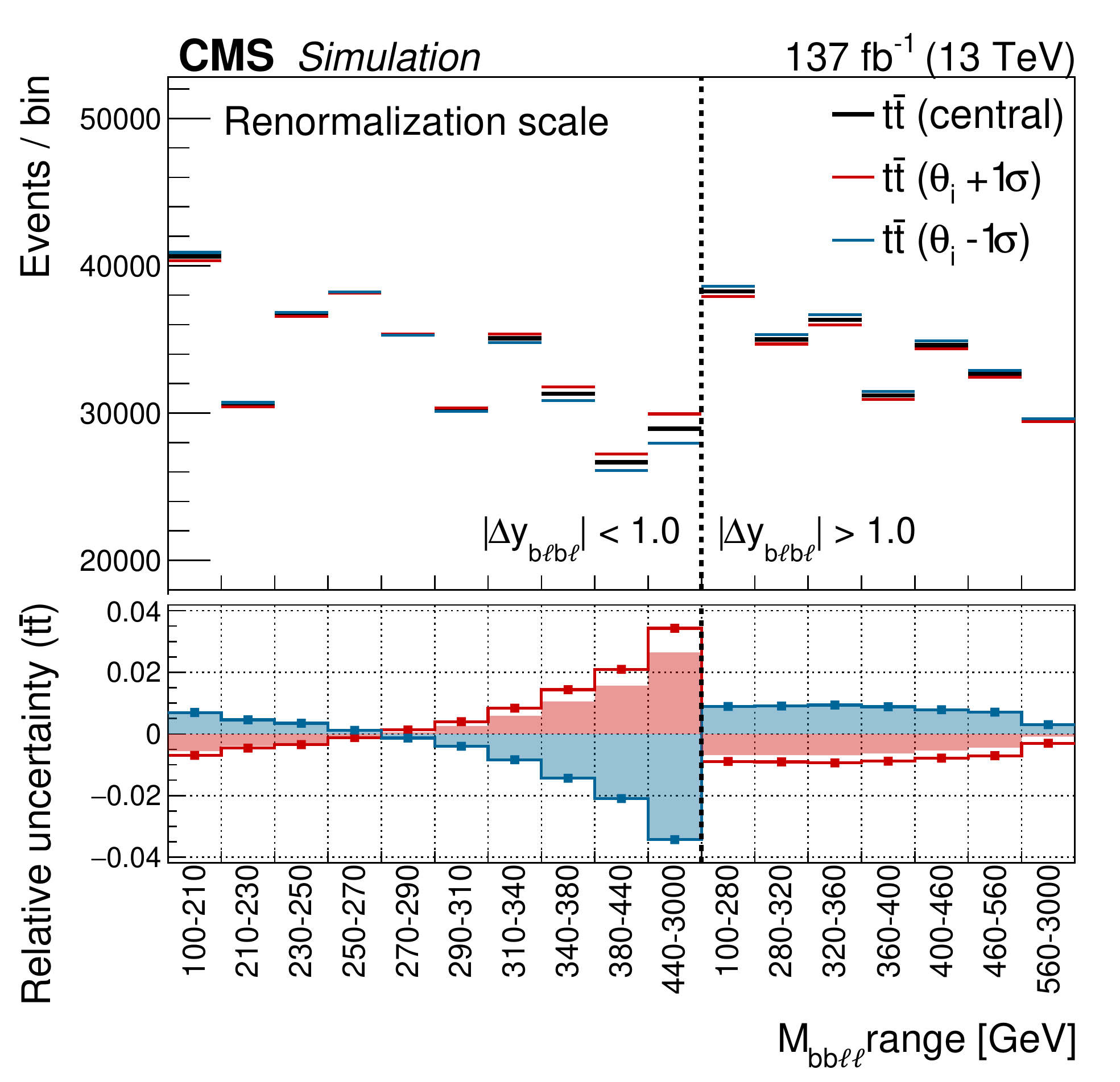}

\caption{Templates are shown for the uncertainties associated with the final-state radiation in \PYTHIA (upper left), the jet energy corrections (upper right), the factorization scale (lower left), and the renormalization scale (lower right). Along with the intrinsic uncertainty in the EW corrections, these are the limiting uncertainties in the fit. The shaded bars represent the raw template information, while the lines show the shapes after smoothing and symmetrization procedures have been applied. In the fit, the jet energy corrections are split into 26 different components, but for brevity only the total uncertainty is shown here. Variation between years is minimal for each of these uncertainties, although they are treated separately in the fit.}
\label{fig:DomUncs}
\end{figure*}

We obtain a fit result of \ytresultfull and an approximate upper limit at 95\% \CL of $\yt<1.54$, where the latter is determined from the point at which $-2\ln(\mathcal{L}(\yt))$ increases by an amount of $1.64^2$ relative to the minimum value. For comparison, the standard model expectation based on simulated Asimov data~\cite{Cowan:2010js__Asimov} is ${\yt}=1\,^{+0.30}_{-0.57}(\text{tot})$ with $\yt<1.47$ at 95\% \CLnp. The scan of the profile likelihood test statistic used to build these intervals is shown in Fig.~\ref{fig:scan}, along with a comparison to the expected behavior based on simulated Asimov data sets. We also show the agreement of data and simulation after performing the fit in Fig.~\ref{fig:postfit}. The minimum of the negative log likelihood occurs at a configuration with good agreement between data and simulation. The result is seen to be clearly limited by systematic uncertainties rather than statistical uncertainty. The templates for the four uncertainties with the greatest effect on the fit are shown in Fig.\,\ref{fig:DomUncs}. 

This result is in agreement with the previously obtained measurement in the lepton+jets final state in Ref.~\cite{ytpaper}, while obtaining a slight increase in sensitivity. Using a different decay channel and a larger data set provides a measurement complementary to the previous result. 

\section{Summary}
\label{S:conc}

A measurement of the Higgs Yukawa coupling to the top quark is presented, based on data from proton-proton collisions collected by the CMS experiment. Data at a center-of-mass energy of 13\TeV is analyzed from the LHC Run 2, collected in 2016--18 and corresponding to an integrated luminosity of 137\fbinv. The resulting best fit value of the top quark Yukawa coupling relative to the standard model is given by \ytresult. This measurement uses the effects of virtual Higgs boson exchange on \ttbar kinematic properties to extract information about the coupling from kinematic distributions. Although the sensitivity is lower compared to constraints obtained from studying processes involving Higgs boson production in Refs.~\cite{ttH} and~\cite{combinedYukawa}, this measurement avoids dependence on other Yukawa coupling values through additional branching assumptions, making it a compelling independent measurement. This measurement also achieves a slightly higher precision than the only other $\yt$ measurement that does not make additional branching fraction assumptions, performed in the search for production of four top quarks. The four top quark search places \yt$ < 1.7$ at a 95\% confidence level~\cite{fourtopRun2} while this measurement achieves an approximate result of \yt $<\twosigmalim$.

\begin{acknowledgments}
We congratulate our colleagues in the CERN accelerator departments for the excellent performance of the LHC and thank the technical and administrative staffs at CERN and at other CMS institutes for their contributions to the success of the CMS effort. In addition, we gratefully acknowledge the computing centers and personnel of the Worldwide LHC Computing Grid for delivering so effectively the computing infrastructure essential to our analyses. Finally, we acknowledge the enduring support for the construction and operation of the LHC and the CMS detector provided by the following funding agencies: BMBWF and FWF (Austria); FNRS and FWO (Belgium); CNPq, CAPES, FAPERJ, FAPERGS, and FAPESP (Brazil); MES (Bulgaria); CERN; CAS, MoST, and NSFC (China); COLCIENCIAS (Colombia); MSES and CSF (Croatia); RIF (Cyprus); SENESCYT (Ecuador); MoER, ERC IUT, PUT and ERDF (Estonia); Academy of Finland, MEC, and HIP (Finland); CEA and CNRS/IN2P3 (France); BMBF, DFG, and HGF (Germany); GSRT (Greece); NKFIA (Hungary); DAE and DST (India); IPM (Iran); SFI (Ireland); INFN (Italy); MSIP and NRF (Republic of Korea); MES (Latvia); LAS (Lithuania); MOE and UM (Malaysia); BUAP, CINVESTAV, CONACYT, LNS, SEP, and UASLP-FAI (Mexico); MOS (Montenegro); MBIE (New Zealand); PAEC (Pakistan); MSHE and NSC (Poland); FCT (Portugal); JINR (Dubna); MON, RosAtom, RAS, RFBR, and NRC KI (Russia); MESTD (Serbia); SEIDI, CPAN, PCTI, and FEDER (Spain); MOSTR (Sri Lanka); Swiss Funding Agencies (Switzerland); MST (Taipei); ThEPCenter, IPST, STAR, and NSTDA (Thailand); TUBITAK and TAEK (Turkey); NASU (Ukraine); STFC (United Kingdom); DOE and NSF (USA).

\hyphenation{Rachada-pisek} Individuals have received support from the Marie-Curie program and the European Research Council and Horizon 2020 Grant, contract Nos.\ 675440, 752730, and 765710 (European Union); the Leventis Foundation; the A.P.\ Sloan Foundation; the Alexander von Humboldt Foundation; the Belgian Federal Science Policy Office; the Fonds pour la Formation \`a la Recherche dans l'Industrie et dans l'Agriculture (FRIA-Belgium); the Agentschap voor Innovatie door Wetenschap en Technologie (IWT-Belgium); the F.R.S.-FNRS and FWO (Belgium) under the ``Excellence of Science -- EOS" -- be.h project n.\ 30820817; the Beijing Municipal Science \& Technology Commission, No. Z191100007219010; the Ministry of Education, Youth and Sports (MEYS) of the Czech Republic; the Deutsche Forschungsgemeinschaft (DFG) under Germany's Excellence Strategy -- EXC 2121 ``Quantum Universe" -- 390833306; the Lend\"ulet (``Momentum") Program and the J\'anos Bolyai Research Scholarship of the Hungarian Academy of Sciences, the New National Excellence Program \'UNKP, the NKFIA research grants 123842, 123959, 124845, 124850, 125105, 128713, 128786, and 129058 (Hungary); the Council of Science and Industrial Research, India; the HOMING PLUS program of the Foundation for Polish Science, cofinanced from European Union, Regional Development Fund, the Mobility Plus program of the Ministry of Science and Higher Education, the National Science Center (Poland), contracts Harmonia 2014/14/M/ST2/00428, Opus 2014/13/B/ST2/02543, 2014/15/B/ST2/03998, and 2015/19/B/ST2/02861, Sonata-bis 2012/07/E/ST2/01406; the National Priorities Research Program by Qatar National Research Fund; the Ministry of Science and Higher Education, project no. 02.a03.21.0005 (Russia); the Tomsk Polytechnic University Competitiveness Enhancement Program; the Programa Estatal de Fomento de la Investigaci{\'o}n Cient{\'i}fica y T{\'e}cnica de Excelencia Mar\'{\i}a de Maeztu, grant MDM-2015-0509 and the Programa Severo Ochoa del Principado de Asturias; the Thalis and Aristeia programs cofinanced by EU-ESF and the Greek NSRF; the Rachadapisek Sompot Fund for Postdoctoral Fellowship, Chulalongkorn University and the Chulalongkorn Academic into Its 2nd Century Project Advancement Project (Thailand); the Graduate Research Fellowship Program of the National Science Foundation, grant no. DGE-141911; the Kavli Foundation; the Nvidia Corporation; the SuperMicro Corporation; the Welch Foundation, contract C-1845; and the Weston Havens Foundation (USA).
\end{acknowledgments}
\bibliography{auto_generated}

\cleardoublepage \appendix\section{The CMS Collaboration \label{app:collab}}\begin{sloppypar}\hyphenpenalty=5000\widowpenalty=500\clubpenalty=5000\input{TOP-19-008-authorlist.tex}\end{sloppypar}
\end{document}

%% file: TOP-19-008-authorlist.tex
\vskip\cmsinstskip
\textbf{Yerevan Physics Institute, Yerevan, Armenia}\\*[0pt]
A.M.~Sirunyan$^{\textrm{\dag}}$, A.~Tumasyan
\vskip\cmsinstskip
\textbf{Institut f\"{u}r Hochenergiephysik, Wien, Austria}\\*[0pt]
W.~Adam, F.~Ambrogi, T.~Bergauer, M.~Dragicevic, J.~Er\"{o}, A.~Escalante~Del~Valle, R.~Fr\"{u}hwirth\cmsAuthorMark{1}, M.~Jeitler\cmsAuthorMark{1}, N.~Krammer, L.~Lechner, D.~Liko, T.~Madlener, I.~Mikulec, F.M.~Pitters, N.~Rad, J.~Schieck\cmsAuthorMark{1}, R.~Sch\"{o}fbeck, M.~Spanring, S.~Templ, W.~Waltenberger, C.-E.~Wulz\cmsAuthorMark{1}, M.~Zarucki
% ...
\vskip\cmsinstskip
\dag: Deceased\\
1:  Also at Vienna University of Technology, Vienna, Austria\\
% ...


%% file: TOP-19-008_temp.bbl
\providecommand{\href}[2]{#2}\begingroup\raggedright\begin{thebibliography}{10}%
\makeatletter
\providecommand{\hrefCMSnoop }[0]{\@secondoftwo}%
\makeatother
\providecommand{\doi}{\texttt{doi:}\begingroup \urlstyle{tt}\Url}

\bibitem{higgsATLAS}
\hrefCMSnoop {}{{ATLAS Collaboration}, ``{Observation of a new particle in the
  search for the standard model Higgs boson with the ATLAS detector at the
  LHC}'',} \textit{ Phys. Lett. B} \textbf{ 716} (2012) 1,
  \href{http://dx.doi.org/10.1016/j.physletb.2012.08.020}{\doi{10.1016/j.physletb.2012.08.020}},
\href{http://www.arXiv.org/abs/1207.7214}{\texttt{arXiv:1207.7214}}.

\bibitem{Chatrchyan:2012xdj_higgsCMSlong}
\hrefCMSnoop {}{{CMS Collaboration}, ``{Observation of a new boson at a mass of
  125 GeV with the CMS experiment at the LHC}'',} \textit{ Phys. Lett. B}
  \textbf{ 716} (2012) 30,
  \href{http://dx.doi.org/10.1016/j.physletb.2012.08.021}{\doi{10.1016/j.physletb.2012.08.021}},
\href{http://www.arXiv.org/abs/1207.7235}{\texttt{arXiv:1207.7235}}.

\bibitem{topmassATLAS}
\hrefCMSnoop {}{{ATLAS Collaboration}, ``{Measurement of the top-quark mass in
  $t\bar{t}+1$-jet events collected with the ATLAS detector in $pp$ collisions
  at $\sqrt{s}=8$ TeV}'',} \textit{ JHEP} \textbf{ 11} (2019) 150,
  \href{http://dx.doi.org/10.1007/JHEP11(2019)150}{\doi{10.1007/JHEP11(2019)150}},
  \href{http://www.arXiv.org/abs/1905.02302}{\texttt{arXiv:1905.02302}}.

\bibitem{Sirunyan:2018gqx_topmass18}
\hrefCMSnoop {}{{CMS Collaboration}, ``{Measurement of the top quark mass with
  lepton+jets final states using $\mathrm {p}$ $\mathrm {p}$ collisions at
  $\sqrt{s}=13\,\text {TeV} $}'',} \textit{ Eur. Phys. J. C} \textbf{ 78}
  (2018) 891,
  \href{http://dx.doi.org/10.1140/epjc/s10052-018-6332-9}{\doi{10.1140/epjc/s10052-018-6332-9}},
\href{http://www.arXiv.org/abs/1805.01428}{\texttt{arXiv:1805.01428}}.

\bibitem{twoHDM}
G.~C. Branco\hrefCMSnoop {}{ {et~al.}, ``{Theory and phenomenology of
  two-Higgs-doublet models}'',} \textit{ Phys. Rept.} \textbf{ 516} (2012) 1,
  \href{http://dx.doi.org/10.1016/j.physrep.2012.02.002}{\doi{10.1016/j.physrep.2012.02.002}},
\href{http://www.arXiv.org/abs/1106.0034}{\texttt{arXiv:1106.0034}}.

\bibitem{minHcomp}
\hrefCMSnoop {}{K.~Agashe, R.~Contino, and A.~Pomarol, ``{The minimal composite
  Higgs model}'',} \textit{ Nucl. Phys. B} \textbf{ 719} (2005) 165,
  \href{http://dx.doi.org/10.1016/j.nuclphysb.2005.04.035}{\doi{10.1016/j.nuclphysb.2005.04.035}},
\href{http://www.arXiv.org/abs/hep-ph/0412089}{\texttt{arXiv:hep-ph/0412089}}.

\bibitem{kappaFramework}
\hrefCMSnoop {}{{LHC Higgs Cross Section Working Group}, ``{Handbook of LHC
  Higgs cross sections: 3. Higgs properties}'',} \textit{ CERN} (2013)
  \href{http://dx.doi.org/10.5170/CERN-2013-004}{\doi{10.5170/CERN-2013-004}},
\href{http://www.arXiv.org/abs/1307.1347}{\texttt{arXiv:1307.1347}}.

\bibitem{Sirunyan:2018lzm_THQ}
\hrefCMSnoop {}{{CMS Collaboration}, ``{Search for associated production of a
  Higgs boson and a single top quark in proton-proton collisions at $\sqrt{s}=
  13$ TeV}'',} \textit{ Phys. Rev. D} \textbf{ 99} (2019) 092005,
  \href{http://dx.doi.org/10.1103/PhysRevD.99.092005}{\doi{10.1103/PhysRevD.99.092005}},
\href{http://www.arXiv.org/abs/1811.09696}{\texttt{arXiv:1811.09696}}.

\bibitem{ttH}
\hrefCMSnoop {}{{CMS Collaboration}, ``{Observation of
  $\mathrm{t\overline{t}}$H production}'',} \textit{ Phys. Rev. Lett.} \textbf{
  120} (2018) 231801,
  \href{http://dx.doi.org/10.1103/PhysRevLett.120.231801}{\doi{10.1103/PhysRevLett.120.231801}},
\href{http://www.arXiv.org/abs/1804.02610}{\texttt{arXiv:1804.02610}}.

\bibitem{tthATLAS}
\hrefCMSnoop {}{{ATLAS Collaboration}, ``{Observation of Higgs boson production
  in association with a top quark pair at the LHC with the ATLAS detector}'',}
  \textit{ Phys. Lett. B} \textbf{ 784} (2018) 173,
  \href{http://dx.doi.org/10.1016/j.physletb.2018.07.035}{\doi{10.1016/j.physletb.2018.07.035}},
\href{http://www.arXiv.org/abs/1806.00425}{\texttt{arXiv:1806.00425}}.

\bibitem{combinedYukawa}
\hrefCMSnoop {}{{CMS Collaboration}, ``{Combined measurements of Higgs boson
  couplings in proton-proton collisions at $\sqrt{s}=$ 13 TeV}'',} \textit{
  Eur. Phys. J. C} \textbf{ 79} (2019) 421,
  \href{http://dx.doi.org/10.1140/epjc/s10052-019-6909-y}{\doi{10.1140/epjc/s10052-019-6909-y}},
\href{http://www.arXiv.org/abs/1809.10733}{\texttt{arXiv:1809.10733}}.

\bibitem{fourtopRun2}
\hrefCMSnoop {}{{CMS Collaboration}, ``{Search for production of four top
  quarks in final states with same-sign or multiple leptons in proton-proton
  collisions at $\sqrt{s}=$ 13 TeV}'',} \textit{ Eur. Phys. J. C} \textbf{ 80}
  (2020) 75,
  \href{http://dx.doi.org/10.1140/epjc/s10052-019-7593-7}{\doi{10.1140/epjc/s10052-019-7593-7}},
\href{http://www.arXiv.org/abs/1908.06463}{\texttt{arXiv:1908.06463}}.

\bibitem{ytpaper}
\hrefCMSnoop {}{{CMS Collaboration}, ``{Measurement of the top quark Yukawa
  coupling from $\mathrm{t\bar{t}}$ kinematic distributions in the lepton+jets
  final state in proton-proton collisions at $\sqrt{s} =$ 13 TeV}'',} \textit{
  Phys. Rev. D} \textbf{ 100} (2019) 072007,
  \href{http://dx.doi.org/10.1103/PhysRevD.100.072007}{\doi{10.1103/PhysRevD.100.072007}},
\href{http://www.arXiv.org/abs/1907.01590}{\texttt{arXiv:1907.01590}}.

\bibitem{UwerWeak}
\hrefCMSnoop {}{J.~H. K{\"u}hn, A.~Scharf, and P.~Uwer, ``Weak interactions in
  top-quark pair production at hadron colliders: an update'',} \textit{ Phys.
  Rev. D} \textbf{ 91} (2015) 014020,
  \href{http://dx.doi.org/10.1103/PhysRevD.91.014020}{\doi{10.1103/PhysRevD.91.014020}},
\href{http://www.arXiv.org/abs/1305.5773}{\texttt{arXiv:1305.5773}}.

\bibitem{hathorart}
M.~Aliev\hrefCMSnoop {}{ {et~al.}, ``{HATHOR: HAdronic Top and Heavy quarks
  crOss section calculatoR}'',} \textit{ Comput. Phys. Commun.} \textbf{ 182}
  (2011) 1034,
  \href{http://dx.doi.org/10.1016/j.cpc.2010.12.040}{\doi{10.1016/j.cpc.2010.12.040}},
\href{http://www.arXiv.org/abs/1007.1327}{\texttt{arXiv:1007.1327}}.

\bibitem{ref:particleflow}
\hrefCMSnoop {}{{CMS Collaboration}, ``Particle-flow reconstruction and global
  event description with the cms detector'',} \textit{ JINST} \textbf{ 12}
  (2017) P10003,
  \href{http://dx.doi.org/10.1088/1748-0221/12/10/P10003}{\doi{10.1088/1748-0221/12/10/P10003}},
\href{http://www.arXiv.org/abs/1706.04965}{\texttt{arXiv:1706.04965}}.

\bibitem{Khachatryan:2016bia}
\hrefCMSnoop {}{{CMS Collaboration}, ``The {CMS} trigger system'',} \textit{
  JINST} \textbf{ 12} (2017) P01020,
  \href{http://dx.doi.org/10.1088/1748-0221/12/01/P01020}{\doi{10.1088/1748-0221/12/01/P01020}},
\href{http://www.arXiv.org/abs/1609.02366}{\texttt{arXiv:1609.02366}}.

\bibitem{Chatrchyan:2008zzk}
\hrefCMSnoop {}{{CMS Collaboration}, ``The {CMS} experiment at the {CERN}
  {LHC}'',} \textit{ JINST} \textbf{ 3} (2008) S08004,
\href{http://dx.doi.org/10.1088/1748-0221/3/08/S08004}{\doi{10.1088/1748-0221/3/08/S08004}}.

\bibitem{Nason:2004rx}
\hrefCMSnoop {}{P.~Nason, ``{A new method for combining NLO QCD with shower
  Monte Carlo algorithms}'',} \textit{ JHEP} \textbf{ 11} (2004) 040,
  \href{http://dx.doi.org/10.1088/1126-6708/2004/11/040}{\doi{10.1088/1126-6708/2004/11/040}},
\href{http://www.arXiv.org/abs/hep-ph/0409146}{\texttt{arXiv:hep-ph/0409146}}.

\bibitem{Frixione:2007vw}
\hrefCMSnoop {}{S.~Frixione, P.~Nason, and C.~Oleari, ``Matching {NLO} {QCD}
  computations with parton shower simulations: the {POWHEG} method'',} \textit{
  JHEP} \textbf{ 11} (2007) 070,
  \href{http://dx.doi.org/10.1088/1126-6708/2007/11/070}{\doi{10.1088/1126-6708/2007/11/070}},
\href{http://www.arXiv.org/abs/0709.2092}{\texttt{arXiv:0709.2092}}.

\bibitem{Alioli:2010xd}
\hrefCMSnoop {}{S.~Alioli, P.~Nason, C.~Oleari, and E.~Re, ``A general
  framework for implementing {NLO} calculations in shower {M}onte {C}arlo
  programs: the {POWHEG BOX}'',} \textit{ JHEP} \textbf{ 06} (2010) 043,
  \href{http://dx.doi.org/10.1007/JHEP06(2010)043}{\doi{10.1007/JHEP06(2010)043}},
\href{http://www.arXiv.org/abs/1002.2581}{\texttt{arXiv:1002.2581}}.

\bibitem{Frixione:2007nw}
\hrefCMSnoop {}{S.~Frixione, P.~Nason, and G.~Ridolfi, ``{A Positive-weight
  next-to-leading-order Monte Carlo for heavy flavour hadroproduction}'',}
  \textit{ JHEP} \textbf{ 09} (2007) 126,
  \href{http://dx.doi.org/10.1088/1126-6708/2007/09/126}{\doi{10.1088/1126-6708/2007/09/126}},
  \href{http://www.arXiv.org/abs/0707.3088}{\texttt{arXiv:0707.3088}}.

\bibitem{Sjostrand:2014zea}
T.~Sj{\"o}strand\hrefCMSnoop {}{ {et~al.}, ``{An introduction to PYTHIA
  8.2}'',} \textit{ Comput. Phys. Commun.} \textbf{ 191} (2015) 159,
  \href{http://dx.doi.org/10.1016/j.cpc.2015.01.024}{\doi{10.1016/j.cpc.2015.01.024}},
\href{http://www.arXiv.org/abs/1410.3012}{\texttt{arXiv:1410.3012}}.

\bibitem{ISR_FSR}
\hrefCMSnoop {}{P.~Skands, S.~Carrazza, and J.~Rojo, ``{Tuning PYTHIA 8.1: the
  Monash 2013 tune}'',} \textit{ Eur. Phys. J. C} \textbf{ 74} (2014) 3024,
  \href{http://dx.doi.org/10.1140/epjc/s10052-014-3024-y}{\doi{10.1140/epjc/s10052-014-3024-y}},
\href{http://www.arXiv.org/abs/1404.5630}{\texttt{arXiv:1404.5630}}.

\bibitem{CP5}
\hrefCMSnoop {}{{CMS Collaboration}, ``{Extraction and validation of a new set
  of CMS PYTHIA8 tunes from underlying-event measurements}'',} \textit{ Eur.
  Phys. J. C} \textbf{ 80} (2020) 4,
  \href{http://dx.doi.org/10.1140/epjc/s10052-019-7499-4}{\doi{10.1140/epjc/s10052-019-7499-4}},
\href{http://www.arXiv.org/abs/1903.12179}{\texttt{arXiv:1903.12179}}.

\bibitem{NNPDF}
\hrefCMSnoop {}{{NNPDF} Collaboration, ``{Parton distributions for the LHC Run
  II}'',} \textit{ JHEP} \textbf{ 04} (2015) 040,
  \href{http://dx.doi.org/10.1007/JHEP04(2015)040}{\doi{10.1007/JHEP04(2015)040}},
\href{http://www.arXiv.org/abs/1410.8849}{\texttt{arXiv:1410.8849}}.

\bibitem{NNPDF31}
\hrefCMSnoop {}{{NNPDF} Collaboration, ``{Parton distributions from
  high-precision collider data}'',} \textit{ Eur. Phys. J. C} \textbf{ 77}
  (2017) 663,
  \href{http://dx.doi.org/10.1140/epjc/s10052-017-5199-5}{\doi{10.1140/epjc/s10052-017-5199-5}},
\href{http://www.arXiv.org/abs/1706.00428}{\texttt{arXiv:1706.00428}}.

\bibitem{Czakon:2011xx}
\hrefCMSnoop {}{M.~Czakon and A.~Mitov, ``Top++: A program for the calculation
  of the top-pair cross-section at hadron colliders'',} \textit{ Comput. Phys.
  Commun.} \textbf{ 185} (2014) 2930,
  \href{http://dx.doi.org/10.1016/j.cpc.2014.06.021}{\doi{10.1016/j.cpc.2014.06.021}},
\href{http://www.arXiv.org/abs/1112.5675}{\texttt{arXiv:1112.5675}}.

\bibitem{Botje:2011sn__pdf4lhc}
\hrefCMSnoop {}{M.~Botje {et~al.}, ``{The PDF4LHC working group interim
  recommendations}'',} (2011).
\href{http://www.arXiv.org/abs/1101.0538}{\texttt{arXiv:1101.0538}}.

\bibitem{Martin:2009iq__9}
\hrefCMSnoop {}{A.~D. Martin, W.~J. Stirling, R.~S. Thorne, and G.~Watt,
  ``{Parton distributions for the LHC}'',} \textit{ Eur. Phys. J. C} \textbf{
  63} (2009) 189,
  \href{http://dx.doi.org/10.1140/epjc/s10052-009-1072-5}{\doi{10.1140/epjc/s10052-009-1072-5}},
\href{http://www.arXiv.org/abs/0901.0002}{\texttt{arXiv:0901.0002}}.

\bibitem{Martin:2009bu__10}
\hrefCMSnoop {}{A.~D. Martin, W.~J. Stirling, R.~S. Thorne, and G.~Watt,
  ``{Uncertainties on \alpS in global PDF analyses and implications for
  predicted hadronic cross sections}'',} \textit{ Eur. Phys. J. C} \textbf{ 64}
  (2009) 653,
  \href{http://dx.doi.org/10.1140/epjc/s10052-009-1164-2}{\doi{10.1140/epjc/s10052-009-1164-2}},
\href{http://www.arXiv.org/abs/0905.3531}{\texttt{arXiv:0905.3531}}.

\bibitem{Lai:2010vv__11}
H.-L. Lai\hrefCMSnoop {}{ {et~al.}, ``{New parton distributions for collider
  physics}'',} \textit{ Phys. Rev. D} \textbf{ 82} (2010) 074024,
  \href{http://dx.doi.org/10.1103/PhysRevD.82.074024}{\doi{10.1103/PhysRevD.82.074024}},
\href{http://www.arXiv.org/abs/1007.2241}{\texttt{arXiv:1007.2241}}.

\bibitem{Gao:2013xoa__12}
J.~Gao\hrefCMSnoop {}{ {et~al.}, ``{CT10 next-to-next-to-leading order global
  analysis of QCD}'',} \textit{ Phys. Rev. D} \textbf{ 89} (2014) 033009,
  \href{http://dx.doi.org/10.1103/PhysRevD.89.033009}{\doi{10.1103/PhysRevD.89.033009}},
\href{http://www.arXiv.org/abs/1302.6246}{\texttt{arXiv:1302.6246}}.

\bibitem{Ball:2012cx}
\hrefCMSnoop {}{{NNPDF} Collaboration, ``{Parton distributions with LHC
  data}'',} \textit{ Nucl. Phys. B} \textbf{ 867} (2013) 244,
  \href{http://dx.doi.org/10.1016/j.nuclphysb.2012.10.003}{\doi{10.1016/j.nuclphysb.2012.10.003}},
\href{http://www.arXiv.org/abs/1207.1303}{\texttt{arXiv:1207.1303}}.

\bibitem{Alwall:2014hca}
J.~Alwall\hrefCMSnoop {}{ {et~al.}, ``{The automated computation of tree-level
  and next-to-leading order differential cross sections, and their matching to
  parton shower simulations}'',} \textit{ JHEP} \textbf{ 07} (2014) 079,
  \href{http://dx.doi.org/10.1007/JHEP07(2014)079}{\doi{10.1007/JHEP07(2014)079}},
\href{http://www.arXiv.org/abs/1405.0301}{\texttt{arXiv:1405.0301}}.

\bibitem{Mangano:2006rw_MLMorig}
\hrefCMSnoop {}{M.~L. Mangano, M.~Moretti, F.~Piccinini, and M.~Treccani,
  ``{Matching matrix elements and shower evolution for top-quark production in
  hadronic collisions}'',} \textit{ JHEP} \textbf{ 01} (2007) 013,
  \href{http://dx.doi.org/10.1088/1126-6708/2007/01/013}{\doi{10.1088/1126-6708/2007/01/013}},
  \href{http://www.arXiv.org/abs/hep-ph/0611129}{\texttt{arXiv:hep-ph/0611129}}.

\bibitem{MLM}
J.~Alwall\hrefCMSnoop {}{ {et~al.}, ``Comparative study of various algorithms
  for the merging of parton showers and matrix elements in hadronic
  collisions'',} \textit{ Eur. Phys. J. C} \textbf{ 53} (2008) 473,
  \href{http://dx.doi.org/10.1140/epjc/s10052-007-0490-5}{\doi{10.1140/epjc/s10052-007-0490-5}},
  \href{http://www.arXiv.org/abs/0706.2569}{\texttt{arXiv:0706.2569}}.

\bibitem{GEANT4}
\hrefCMSnoop {}{{GEANT4} Collaboration, ``{\GEANTfour}---a simulation
  toolkit'',} \textit{ Nucl. Instrum. Meth. A} \textbf{ 506} (2003) 250,
\href{http://dx.doi.org/10.1016/S0168-9002(03)01368-8}{\doi{10.1016/S0168-9002(03)01368-8}}.

\bibitem{ignorephotons}
\hrefCMSnoop {}{W.~Hollik and M.~Kollar, ``{NLO QED contributions to top-pair
  production at hadron collider}'',} \textit{ Phys. Rev. D} \textbf{ 77} (2008)
  014008,
  \href{http://dx.doi.org/10.1103/PhysRevD.77.014008}{\doi{10.1103/PhysRevD.77.014008}},
\href{http://www.arXiv.org/abs/0708.1697}{\texttt{arXiv:0708.1697}}.

\bibitem{Czakon:2017NLOEW}
M.~Czakon\hrefCMSnoop {}{ {et~al.}, ``{Top-pair production at the LHC through
  NNLO QCD and NLO EW}'',} \textit{ JHEP} \textbf{ 10} (2017) 186,
  \href{http://dx.doi.org/10.1007/JHEP10(2017)186}{\doi{10.1007/JHEP10(2017)186}},
\href{http://www.arXiv.org/abs/1705.04105}{\texttt{arXiv:1705.04105}}.

\bibitem{Otto:17002}
\hrefCMSnoop {}{{CMS Collaboration}, ``{Measurement of differential cross
  sections for the production of top quark pairs and of additional jets in
  lepton+jets events from pp collisions at $\sqrt{s} =$ 13 TeV}'',} \textit{
  Phys. Rev. D} \textbf{ 97} (2018) 112003,
  \href{http://dx.doi.org/10.1103/PhysRevD.97.112003}{\doi{10.1103/PhysRevD.97.112003}},
\href{http://www.arXiv.org/abs/1803.08856}{\texttt{arXiv:1803.08856}}.

\bibitem{antikt}
\hrefCMSnoop {}{M.~Cacciari, G.~P. Salam, and G.~Soyez, ``{The anti-\kt jet
  clustering algorithm}'',} \textit{ JHEP} \textbf{ 04} (2008) 063,
  \href{http://dx.doi.org/10.1088/1126-6708/2008/04/063}{\doi{10.1088/1126-6708/2008/04/063}},
\href{http://www.arXiv.org/abs/0802.1189}{\texttt{arXiv:0802.1189}}.

\bibitem{fastjet}
\hrefCMSnoop {}{M.~Cacciari, G.~P. Salam, and G.~Soyez, ``{FastJet user
  manual}'',} \textit{ Eur. Phys. J. C} \textbf{ 72} (2012) 1896,
  \href{http://dx.doi.org/10.1140/epjc/s10052-012-1896-2}{\doi{10.1140/epjc/s10052-012-1896-2}},
\href{http://www.arXiv.org/abs/1111.6097}{\texttt{arXiv:1111.6097}}.

\bibitem{JES}
\hrefCMSnoop {}{{CMS Collaboration}, ``{Jet energy scale and resolution in the
  CMS experiment in pp collisions at 8 TeV}'',} \textit{ JINST} \textbf{ 12}
  (2017) P02014,
  \href{http://dx.doi.org/10.1088/1748-0221/12/02/P02014}{\doi{10.1088/1748-0221/12/02/P02014}},
\href{http://www.arXiv.org/abs/1607.03663}{\texttt{arXiv:1607.03663}}.

\bibitem{btagRun2}
\hrefCMSnoop {}{{CMS Collaboration}, ``{Identification of heavy-flavour jets
  with the CMS detector in pp collisions at 13 TeV}'',} \textit{ JINST}
  \textbf{ 13} (2018) P05011,
  \href{http://dx.doi.org/10.1088/1748-0221/13/05/P05011}{\doi{10.1088/1748-0221/13/05/P05011}},
\href{http://www.arXiv.org/abs/1712.07158}{\texttt{arXiv:1712.07158}}.

\bibitem{burt}
\hrefCMSnoop {}{B.~A. Betchart, R.~Demina, and A.~Harel, ``Analytic solutions
  for neutrino momenta in decay of top quarks'',} \textit{ Nucl. Instrum. Meth.
  A} \textbf{ 736} (2014) 169,
  \href{http://dx.doi.org/10.1016/j.nima.2013.10.039}{\doi{10.1016/j.nima.2013.10.039}},
\href{http://www.arXiv.org/abs/1305.1878}{\texttt{arXiv:1305.1878}}.

\bibitem{Khachatryan:2014jba__LLscan}
\hrefCMSnoop {}{{CMS Collaboration}, ``Precise determination of the mass of the
  {Higgs} boson and tests of compatibility of its couplings with the standard
  model predictions using proton collisions at 7 and 8 {TeV}'',} \textit{ Eur.
  Phys. J. C} \textbf{ 75} (2015) 212,
  \href{http://dx.doi.org/10.1140/epjc/s10052-015-3351-7}{\doi{10.1140/epjc/s10052-015-3351-7}},
\href{http://www.arXiv.org/abs/1412.8662}{\texttt{arXiv:1412.8662}}.

\bibitem{lumi16}
\href {https://cds.cern.ch/record/2257069}{{CMS Collaboration}, ``{CMS
  Luminosity Measurements for the 2016 Data Taking Period}'',} CMS Physics
  Analysis Summary CMS-PAS-LUM-17-001, 2017.

\bibitem{lumi17}
\href {https://cds.cern.ch/record/2621960}{{CMS Collaboration}, ``{CMS
  luminosity measurement for the 2017 data-taking period at $\sqrt{s} =
  13~\mathrm{TeV}$}'',} CMS Physics Analysis Summary CMS-PAS-LUM-17-004, 2018.

\bibitem{lumi18}
\href {https://cds.cern.ch/record/2676164}{{CMS Collaboration}, ``{CMS
  luminosity measurement for the 2018 data-taking period at $\sqrt{s} =
  13~\mathrm{TeV}$}'',} CMS Physics Analysis Summary CMS-PAS-LUM-18-002, 2019.

\bibitem{InelasticXS}
\hrefCMSnoop {}{{ATLAS Collaboration}, ``{Measurement of the inelastic
  proton-proton cross section at $\sqrt{s} = 13$\TeV with the ATLAS detector at
  the LHC}'',} \textit{ Phys. Rev. Lett.} \textbf{ 117} (2016) 182002,
  \href{http://dx.doi.org/10.1103/PhysRevLett.117.182002}{\doi{10.1103/PhysRevLett.117.182002}},
\href{http://www.arXiv.org/abs/1606.02625}{\texttt{arXiv:1606.02625}}.

\bibitem{Khachatryan:2010xn__DYtnp}
\hrefCMSnoop {}{{CMS Collaboration}, ``{Measurements of Inclusive $W$ and $Z$
  Cross Sections in pp Collisions at $\sqrt{s}=7$ TeV}'',} \textit{ JHEP}
  \textbf{ 01} (2011) 080,
  \href{http://dx.doi.org/10.1007/JHEP01(2011)080}{\doi{10.1007/JHEP01(2011)080}},
\href{http://www.arXiv.org/abs/1012.2466}{\texttt{arXiv:1012.2466}}.

\bibitem{electronPerformance8TeV}
\hrefCMSnoop {}{{CMS Collaboration}, ``{Performance of electron reconstruction
  and selection with the CMS detector in proton-proton collisions at $\sqrt{s}
  = 8$\TeV}'',} \textit{ JINST} \textbf{ 10} (2015) P06005,
  \href{http://dx.doi.org/10.1088/1748-0221/10/06/P06005}{\doi{10.1088/1748-0221/10/06/P06005}},
\href{http://www.arXiv.org/abs/1502.02701}{\texttt{arXiv:1502.02701}}.

\bibitem{muonPerformance13TeV}
\hrefCMSnoop {}{{CMS Collaboration}, ``{Performance of the CMS muon detector
  and muon reconstruction with proton-proton collisions at $\sqrt{s}=$ 13
  TeV}'',} \textit{ JINST} \textbf{ 13} (2018) P06015,
  \href{http://dx.doi.org/10.1088/1748-0221/13/06/P06015}{\doi{10.1088/1748-0221/13/06/P06015}},
\href{http://www.arXiv.org/abs/1804.04528}{\texttt{arXiv:1804.04528}}.

\bibitem{PDG}
\hrefCMSnoop {}{{Particle Data Group}, M.~Tanabashi {et~al.}, ``Review of
  particle physics'',} \textit{ Phys. Rev. D} \textbf{ 98} (2018) 030001,
  \href{http://dx.doi.org/10.1103/PhysRevD.98.030001}{\doi{10.1103/PhysRevD.98.030001}}.

\bibitem{Bowler}
\hrefCMSnoop {}{M.~G. Bowler, ``{${\rm e}^{+}{\rm e}^{-}$ Production of heavy
  quarks in the string model}'',} \textit{ Z. Phys. C} \textbf{ 11} (1981) 169,
\href{http://dx.doi.org/10.1007/BF01574001}{\doi{10.1007/BF01574001}}.

\bibitem{topPTexample}
\hrefCMSnoop {}{{CMS Collaboration}, ``{Measurement of the
  $\mathrm{t}\overline{\mathrm{t}}$ production cross section, the top quark
  mass, and the strong coupling constant using dilepton events in pp collisions
  at $\sqrt{s} =$ 13 TeV}'',} \textit{ Eur. Phys. J. C} \textbf{ 79} (2019)
  368,
  \href{http://dx.doi.org/10.1140/epjc/s10052-019-6863-8}{\doi{10.1140/epjc/s10052-019-6863-8}},
  \href{http://www.arXiv.org/abs/1812.10505}{\texttt{arXiv:1812.10505}}.

\bibitem{Grazzini:2017mhc__matrix}
\hrefCMSnoop {}{M.~Grazzini, S.~Kallweit, and M.~Wiesemann, ``{Fully
  differential NNLO computations with MATRIX}'',} \textit{ Eur. Phys. J. C}
  \textbf{ 78} (2018) 537,
  \href{http://dx.doi.org/10.1140/epjc/s10052-018-5771-7}{\doi{10.1140/epjc/s10052-018-5771-7}},
\href{http://www.arXiv.org/abs/1711.06631}{\texttt{arXiv:1711.06631}}.

\bibitem{Czakon:fnlo1}
\hrefCMSnoop {}{M.~Czakon, D.~Heymes, and A.~Mitov, ``{\textsc{fastNLO} tables
  for NNLO top-quark pair differential distributions}'',} (2017).
\href{http://www.arXiv.org/abs/1704.08551}{\texttt{arXiv:1704.08551}}.

\bibitem{Czakon:fnlo2}
\hrefCMSnoop {}{M.~Czakon, D.~Heymes, and A.~Mitov, ``{High-precision
  differential predictions for top-quark pairs at the LHC}'',} \textit{ Phys.
  Rev. Lett.} \textbf{ 116} (2016) 082003,
  \href{http://dx.doi.org/10.1103/PhysRevLett.116.082003}{\doi{10.1103/PhysRevLett.116.082003}},
\href{http://www.arXiv.org/abs/1511.00549}{\texttt{arXiv:1511.00549}}.

\bibitem{Czakon:2Dfnlo}
M.~Czakon\hrefCMSnoop {}{ {et~al.}, ``{A study of the impact of
  double-differential top distributions from CMS on parton distribution
  functions}'',} (2019).
\href{http://www.arXiv.org/abs/1912.08801}{\texttt{arXiv:1912.08801}}.

\bibitem{LOWESS}
\hrefCMSnoop {}{W.~S. Cleveland, ``Robust locally weighted regression and
  smoothing scatterplots'',} \textit{ J. Am. Stat. Assoc.} \textbf{ 74} (1979)
  829,
  \href{http://dx.doi.org/10.1080/01621459.1979.10481038}{\doi{10.1080/01621459.1979.10481038}}.

\bibitem{Cowan:2010js__Asimov}
\hrefCMSnoop {}{G.~Cowan, K.~Cranmer, E.~Gross, and O.~Vitells, ``Asymptotic
  formulae for likelihood-based tests of new physics'',} \textit{ Eur. Phys. J.
  C} \textbf{ 71} (2011) 1554,
  \href{http://dx.doi.org/10.1140/epjc/s10052-011-1554-0}{\doi{10.1140/epjc/s10052-011-1554-0}},
  \href{http://www.arXiv.org/abs/1007.1727}{\texttt{arXiv:1007.1727}}.
[Erratum: \DOI{10.1140/epjc/s10052-013-2501-z}].

\end{thebibliography}\endgroup
